%% file: ms.tex
\documentclass[12pt,letterpaper]{article}
\input{Header.tex}

\date{July 28, 2024}
\title{Optimal Pre-Analysis Plans:\\
Statistical Decisions Subject to Implementability}
\author{Maximilian Kasy\footnote{Department of Economics, University of Oxford. \href{mailto:maximilian.kasy@economics.ox.ac.uk}{maximilian.kasy@economics.ox.ac.uk}.} \and Jann Spiess\footnote{Graduate School of Business, Stanford University. \href{mailto:jspiess@stanford.edu}{jspiess@stanford.edu}.\vspace{6pt}}}

\begin{document}

\maketitle

\onehalfspacing
\vspace{-6pt}
\begin{abstract}
\input{Sections/Abstract}  
\end{abstract}

\extrafootertext{
  We thank Stefano DellaVigna, Ted Miguel, Marco Ottaviani, and Davide Viviano, as well as
  Alex Frankel, Carlos Gonzalez Perez, Rohit Lamba, Ludvig Sinander, Alex Teytelboym, and participants at the BITSS 2022 meeting, the 2022 AEA meetings, and the 2022 conference in Honor of Jim Powell for helpful discussions and suggestions. \\
  Maximilian Kasy was supported by the Alfred P. Sloan Foundation, under the grant ``Social foundations for statistics and machine learning.''
}

\clearpage
\input{Sections/Introduction.tex}
\input{Sections/Example.tex}
\input{Sections/Model.tex}

\input{Sections/Implementability.tex}

\input{Sections/Testing.tex}

\input{Sections/Case_studies.tex}

\input{Sections/Conclusion.tex}

\clearpage
\appendix

\input{Sections/Proofs.tex}

\clearpage
\input{Sections/Simple.tex}

\clearpage
\bibliographystyle{apalikefull}
\bibliography{Literature}

\end{document}

%% file: Header.tex
\usepackage[utf8]{inputenc}
\usepackage{amsmath}
\usepackage{amsfonts}
\usepackage{amssymb}
\usepackage{amsthm}
\usepackage{mathrsfs}
\usepackage{natbib}
\usepackage{todonotes}
\usepackage{graphicx}
\usepackage{soul}
\usepackage{comment}
\usepackage{xspace}
\usepackage{mdframed}
\usepackage{setspace}
\usepackage{versions}
\usepackage[export]{adjustbox}
\usepackage{booktabs}
\usepackage{mathtools}
\usepackage{adjustbox}
\usepackage{placeins}
\usepackage{subcaption}

\usetikzlibrary{positioning, arrows.meta}

\newcommand{\M}{\mathcal{M}}
\newcommand{\bs}{\boldsymbol}
\def\bal#1\eal{\begin{align}#1\end{align}}
\def\bals#1\eals{\begin{align*}#1\end{align*}}
\def\be#1\ee{\begin{equation}#1\end{equation}}

\newcommand{\psm}[1]{\begin{psmallmatrix}#1\end{psmallmatrix}}

\newcommand{\act}{\mathbf{a}}

\newcommand{\Act}{A}
\newcommand{\actreduced}{\bar{\act}}

\newcommand{\param}{\theta}

\newcommand\extrafootertext[1]{%
    \bgroup
    \renewcommand\thefootnote{\fnsymbol{footnote}}%
    \renewcommand\thempfootnote{\fnsymbol{mpfootnote}}%
    \footnotetext[0]{#1}%
    \egroup
}

\newcommand{\R}{\mathbb{R}}
\let\P\relax
\DeclareMathOperator{\P}{P}
\DeclareMathOperator{\E}{E}
\newcommand{\X}{\mathcal{X}}

\DeclareMathOperator*{\argmax}{argmax\;}

\newcommand{\N}{\mathcal{N}}

\newcommand{\B}{\mathcal{B}}

\makeatletter%
\@ifclassloaded{beamer}%
  {

  }%
  {
    \usepackage[breaklinks=true,hidelinks]{hyperref}
    \hypersetup{
        colorlinks,
        linkcolor={red!50!black},
        citecolor={blue!50!black},
        urlcolor={blue!80!black}
    }

    \usepackage[margin=1.25in]{geometry}
    \usepackage[shortlabels]{enumitem}

    \newtheorem{definition}{Definition}
    \newtheorem{theorem}{Theorem}
    
    \newtheorem{lemma}{Lemma}
    
    \newtheorem{assumption}{Assumption}
  }%
\makeatother%
    
    \newtheorem{proposition}{Proposition}

\newenvironment{proofwithspace}[1][\proofname]{\begin{proof}[#1]\mbox{}\\*}{\end{proof}\vspace{1em}}

\defcitealias{fda1998}{FDA}

%% file: Sections/Abstract.tex
What is the purpose of pre-analysis plans, and how should they be designed?
We model the interaction between an agent who analyzes data
and a principal who makes a decision based on agent reports.
The agent could be the manufacturer of a new drug, and the principal a regulator deciding whether the drug is approved.
Or the agent could be a researcher submitting a research paper, and the principal an editor deciding whether it is published.
The agent decides which statistics to report to the principal.
The principal cannot verify whether the analyst reported selectively. 
Absent a pre-analysis message, if there are conflicts of interest, then many desirable decision rules cannot be implemented.
Allowing the agent to send a message before seeing the data increases the set of decision rules that can be implemented, and allows the principal to leverage agent expertise.
The optimal mechanisms that we characterize require pre-analysis plans.
Applying these results to hypothesis testing, we show that optimal rejection rules pre-register a valid test, and make worst-case assumptions about unreported statistics.
Optimal tests can be found as a solution to a linear-programming problem.

%% file: Sections/Introduction.tex
\section{Introduction}

When writing up their studies, empirical researchers might cherry-pick the findings that they report.
Cherry-picking distorts the inferences that we can draw from published findings.
As a potential solution, 
pre-analysis plans (PAPs) have become a precondition for the publication of experimental research in economics, for both field experiments and lab experiments.%
\footnote{Just as in the case of randomized experiments, the adoption of PAPs in economics follows their prior adoption in clinical research; see for instance the guidelines of the \citetalias{fda1998} on PAPs, \citep{fda1998}.}
PAPs can enable valid inference by pre-specifying a mapping from the data to testing decisions or estimates, cf. \cite{christensen2018transparency,miguel2021evidence}. 
This can prevent the cherry-picking of results, and thus provide a remedy for the distortions introduced by unacknowledged multiple hypothesis testing.
The widespread adoption of PAPs has not gone uncontested, however,%
\footnote{See for instance \cite{CoffmanNiederle2015}, \cite{Olken2015}, and \cite{duflo2020praise}, who discuss the costs and benefits of PAPs in experimental economics from a practitioners' perspective.}
and has been criticized for constraining our ability to learn from experiments.

In this article, we clarify the benefits and optimal design of pre-analysis plans by modeling statistical inference as a mechanism-design problem \citep{myerson1986multistage,kamenica2019bayesian}.
To motivate this approach, note that, in single-agent statistical decision theory, rational decision-makers with preferences that are consistent over time do not need the commitment device that is provided by a PAP.
This holds in particular when a single decision-maker aims to construct tests that control size, or estimators that are unbiased. Single decision-makers have no reason to ``cheat themselves.''
The situation is different, however, when there are multiple agents with conflicting interests. When there are multiple agents, not all statistical decision rules might be implementable. Furthermore, allowing for messages (PAPs) before the data are seen can increase the set of implementable rules, and thus improve welfare.\footnote{A separate argument for pre-analysis plans, which we do not pursue in this paper, might be based on dynamic inconsistencies in agent preferences, for instance because of present-bias.}

Our framework provides a theoretical justification of PAPs. 
In addition to our theoretical results, which are based on this framework, we also derive guidance for practitioners, including both decision-makers (e.g., readers, editors) and data analysts (e.g., study authors). From the decision-makers' perspective, we describe how tests, estimators, or other decision rules can be implemented by requiring pre-analysis plans.
We then focus on hypothesis tests, and describe how to derive optimal pre-analysis plans from the analysts' perspective.
These pre-analysis plans maximize power while controlling size and maintaining implementability.
We furthermore provide software (an interactive web app) to facilitate the design of optimal pre-analysis plans.

\paragraph{Examples}

In our model, we consider the interaction between a decision-maker and an analyst. The analyst has private information and interests which differ from those of the decision-maker.
One example of such a conflict of interest is between a researcher (analyst) who wants to reject a hypothesis, and a reader of their research (decision-maker) who wants a valid statistical test of that same hypothesis; the relevant decision here is whether to reject the null hypothesis.
Another example is the conflict of interest between a researcher (analyst) who wants to get published, and a journal editor (decision-maker) who only wants to publish studies on effects that are large enough to be interesting; the relevant decision here is whether to publish a study.
A third example is the conflict of interest between a pharmaceutical company (analyst) who wants to sell drugs, and a medical regulatory agency (decision-maker) who wants to protect patient health; the relevant decision here is whether to approve a drug.

\paragraph{Model and timeline}

The timeline of our model is as follows.
Before observing the data, the analyst can send a message to the decision-maker. This message might for instance be in the form of a pre-analysis plan.
Then the analyst observes the data. The data are given in the form of a set of statistics, such as the outcomes of different hypothesis tests, or estimates for different model specifications.
The analyst chooses a subset of these statistics to report to the decision-maker.

The decision-maker observes the pre-analysis message and the statistics which the analyst reported, and makes a decision based on this information.
We assume that this decision is real-valued, and that the analyst always prefers a higher value for this decision. We consider different objectives for the decision-maker, including statistical testing subject to size control.

In our model, the analyst can \emph{hide} information from the decision-maker, by not reporting some statistics, but they cannot \emph{lie} about the data that they report. 
The potential value of a pre-analysis message in this model comes from the fact that it allows the analyst to share private information (i.e., expertise) with the decision-maker.
Sharing such information truthfully would not be incentive-compatible if the message could only be sent after seeing the data.
The analyst might have private information regarding the availability of statistics, and regarding the state of the world.

To make it possible for the analyst to hide information, they need to have plausible deniability: The decision-maker does not know what statistics the analyst got to see. Experiments might not have been run, or data might not have been collected, for instance.
The analyst might also have prior uncertainty over the availability of statistics, but this is not necessary for our conclusions.

The mechanism-design approach which motivates our model takes the perspective of a decision-maker who wants to implement a statistical decision rule. Not all rules are implementable, however, when the analyst has divergent interests and private information. This mechanism-design perspective allows us to stay close to standard statistical theory, while taking into account the implementability constraints that are a consequence of the social nature of research.

\paragraph{Implementable decision rules}

For this model, we first characterize the set of implementable statistical decision rules.
This set is independent of decision-maker preferences.
We show that implementable decision rules are such that reporting more results can never make the analyst worse off, given the pre-analysis message, and given the realization of the data. 
Formally, implementable decision rules need to be \emph{monotonic in the reported set} of statistics, in terms of set inclusion. 

Implementable decision rules furthermore need to be compatible with \emph{truthful revelation of analyst private information} prior to observing any data \citep{myerson1986multistage}.
This condition is equivalent to the conditions satisfied by \emph{proper scoring} rules \citep{savage1971elicitation, gneiting2007strictly}.

Pre-analysis messages allow the decision-maker to implement a larger set of decision rules than would be available without such messages.
Implementable rules can be implemented using different mechanisms, based on such pre-analysis messages.
One possible implementation allows the analyst to \emph{choose from a restricted set of decision rules} before seeing the data. Each of these rules needs to be monotonic in the set of reported statistics.
This implementation corresponds to the actual practice of pre-analysis plans, where the analyst chooses a decision rule before the data becomes available.

The set of implementable rules can be characterized as a \emph{convex polytope}. If the decision-maker's objective is convex, and in particular if it is linear, then the optimal implementable rule is necessarily an \emph{extremal point} of this polytope \citep{vanderbei2020linear}.

\paragraph{Optimal implementable hypothesis tests}

We next turn to the specific problem of finding optimal implementable hypothesis tests.
Such tests are required to satisfy \emph{size control} conditional on the state of the world and conditional on analyst private information that is available before observing the data.
We show that the optimal implementable test, for the decision-maker, can be implemented by (i) requiring the analyst to choose an arbitrary \emph{full-data} test, which is a function of all statistics that the analyst might observe, where this test controls size, and then (ii) implementing this test, making \emph{worst-case assumptions} about any unreported statistics.

The analyst's problem of finding a full-data test that maximizes expected power for this mechanism can be cast as a linear programming problem.
If the analyst knows the set of available statistics at the time of writing their pre-analysis plan, this problem reduces to the classic problem of finding a test (based on the full set of available statistics) with high expected power, subject to size control. The solution to this problem takes the form of a likelihood ratio test. 
More generally, the set of available statistics might not be known for sure at the time of writing the PAP. 
We provide an interactive app that allows the analyst to solve the linear programming problem for this case, based on their prior beliefs. The output of our app can serve as a basis for their pre-analysis plan.

\paragraph{Roadmap}

The rest of this article is structured as follows.
We conclude this introduction with a review of some related literature.
In \autoref{sec:motivating_example}, we present a motivating example concerning statistical testing and p-hacking.
In \autoref{sec:model}, we introduce the general model.
In \autoref{sec:implementability}, we characterize implementable decision rules.
In \autoref{sec:testing}, we characterize optimal implementable hypothesis tests.
In \autoref{sec:casestudies}, we illustrate our results by applying them to the setting of \cite{dellavigna2018motivates}, using expert forecasts to construct a prior distribution.
In \autoref{sec:conclusion}, we summarize and discuss some limitations of our model.
\autoref{sec:proofs} contains all proofs.

\subsection{Related literature}

Our article speaks to the current debates around pre-registration -- and other possible reforms -- in empirical economics and other social- and life-sciences; cf. \cite{christensen2018transparency, miguel2021evidence}, which are motivated by the distortions to statistical inference that might be induced by selective reporting, cf. \cite{publicationbias2019, andrews2019inference}.
In doing so, our article applies some of the insights from mechanism design and information design \citep{myerson1986multistage,kamenica2019bayesian,mechanismdesignnotes2023} to the settings of statistical decision theory and statistical testing, \citep{wald1950statistical,savage1951theory,lehmann2006testing}.

More broadly, our article contributes to a literature that spans statistics, econometrics and economic theory, and which models statistical inference in multi-agent settings.
We differ from other contributions to this literature, in that we focus on the role of implementability as a constraint on statistical decision rules, which rationalizes pre-analysis plans, and on the derivation of optimal decision rules subject to the constraint of implementability.

Drawing on classic references \citep{Tullock1959-be, Sterling1959-nu, Leamer1974-zl}, \cite{Glaeser2006-yp} considers the role of incentives in empirical research.
A number of recent contributions model estimation and testing within multiple-agent settings, including \cite{glazer2004optimal, mathis2008full, Chassang2012-is,Tetenov2016-pw,Ottaviani2017-qq, Di_Tillio2017-ur,Spiess2018-nx,Henry2019-pt,McCloskey2020-pu,Libgober2020-ts,Yoder2020-fe,Williams2021-hg,Abrams2021-hz,Viviano2021-wt}.
In  this literature, \cite{Banerjee2020-ql,Frankel_undated-xn, Andrews2020-cp, gao2022inference} consider the communication of scientific results to an audience with priors, information, or objectives that might differ from the sender's.

The literature on  Bayesian persuasion \citep{kamenica2011bayesian,kamenica2019bayesian, curello2022comparative}, like the present article, considers a sender with information unavailable to a receiver, where sender and receiver have divergent objectives.
One important way in which our model differs from that of Bayesian persuasion is that in our model the signal space of the analyst is restricted to the truthful but selective reporting of data. This restriction implies that the concavification argument central to Bayesian persuasion does not apply.%

%% file: Sections/Example.tex
\section{A motivating example}
\label{sec:motivating_example}

Before we introduce our general model, consider the following hypothesis-testing problem, as a motivating example and special case.
The full data consists of two normally distributed statistics, $X = (X_1,X_2)$, with $X_i \sim \mathcal{N}(\param,1)$, independently across components of the vector $X$.
The $X_i$ might for instance correspond to experimental estimates of an average treatment effect, for two different experimental sites.
There is a decision-maker and an analyst.
The decision-maker wants to test the null hypothesis $H_0: \param \leq 0$.
The analyst, however, aims to simply maximize the probability of rejection.

The analyst might not always observe both statistics $X_1,X_2$.
They instead observe the subvector $X_J$ for a random index set $J$.
The possible values of the index set $J$ are $\emptyset$, $\{1\}$, $\{2\}$, and $\{1,2\}$. 
The statistic $X_i$, for $i \in \{1,2\}$, is observed with probability $P(i \in J)$. 
Observability is independent across statistics.
$P(i \in J)$ is the decision-maker's a-priori probability that the analyst successfully implemented an experiment at site $i$.

The decision-maker does not know which statistics are actually available, that is, they do not know $J$. The analyst knows which statistics are available. This allows the analyst to selectively report (``p-hack''), with plausible deniability, since they might not have observed some unreported statistic.
Upon learning the data $X_J$, the analyst chooses a subset $I\subseteq J$, and reports $(X_I,I)$ to the decision-maker. The decision-maker then rejects the null with probability $\act(X_I,I) \in [0,1]$.
How should the decision-maker choose the testing rule $\act$ that maps the reported data to a rejection probability?

\begin{figure}[p]
    \caption{Rejection probabilities for different testing rules}
    \label{fig:testing_rules}
    \vspace{-20pt}
    \includegraphics[width=1.2\textwidth,center]{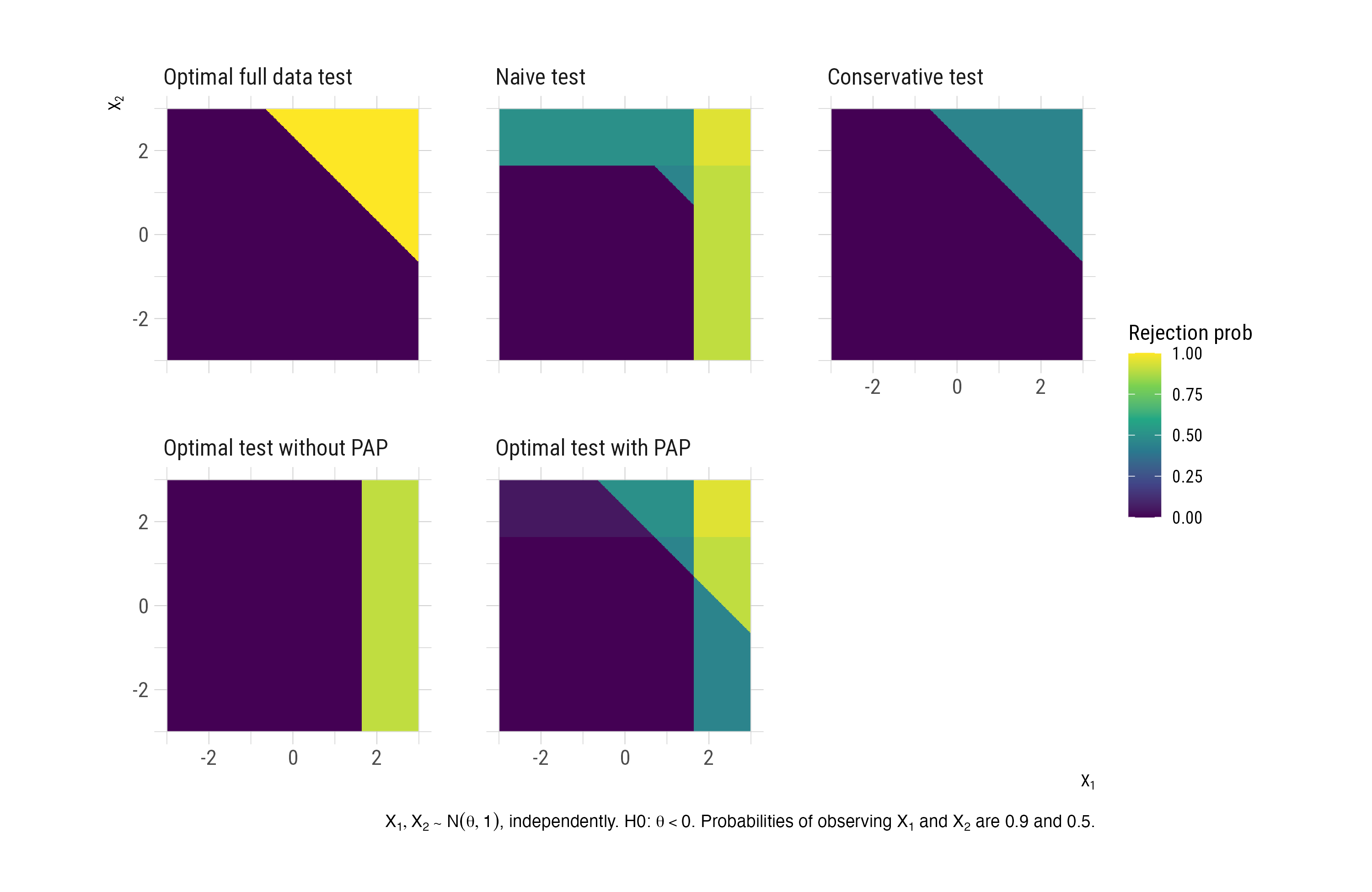}\\
    \caption{Power curves}
    \vspace{-20pt}
    \label{fig:power_curves}
    \vspace{-10pt}
\includegraphics[width=1.2\textwidth,center]{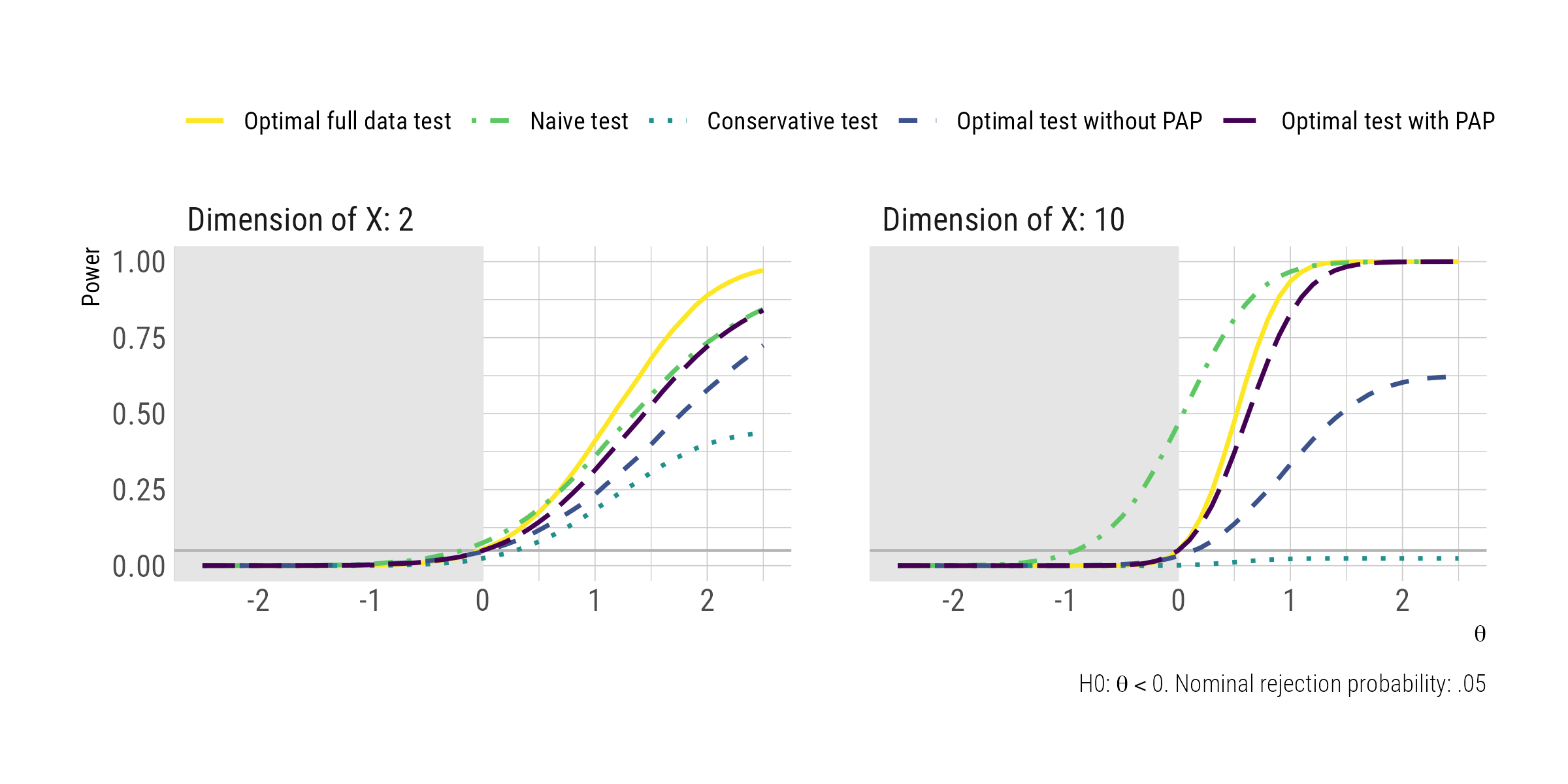}
\end{figure}
    
\paragraph{Five testing rules}
We compare five different testing rules, $\act_1$ through $\act_5$. 
For each of these testing rules, \autoref{fig:testing_rules} shows the rejection probability as a function of $(X_1,X_2)$, assuming that $P(1 \in J) = 0.9$ and $P(2 \in J) = 0.5$.
The rejection probability in \autoref{fig:testing_rules} conditions on $X$, but averages over the distribution of $J$, and takes into account the analyst's endogenous response to a given testing rule.
The left panel of \autoref{fig:power_curves} shows the corresponding power curves, i.e., the rejection probability as a function of $\param$, averaging over the distribution of both $X$ and $J$.

Our benchmark is the \textbf{optimal test using all the data}.
This test is not, in general, feasible, since not all statistics are always available.
We have that $Z = \tfrac{1}{\sqrt{2}}(X_1 + X_2) \sim \N({\sqrt{2}} \cdot \param,1)$ is a sufficient statistic for $\param$. Since this statistic satisfies the monotone likelihood ratio property, the Neyman--Pearson Lemma implies that the uniformly most powerful test of level $\alpha$ is given by 
$\act_1(X) = \bs 1 (Z > z),$
where $z = \Phi^{-1}(1-\alpha)$; cf. Theorem 3.4.1 in \cite{lehmann2006testing}.

Consider next the \textbf{naive test} which ignores potentially selective reporting by the analyst.
This test acts as if the reported statistics $I$ are the full data available to the analyst, and implements the corresponding uniformly most powerful test,
$$
    \act_2(X_I, I) = \bs 1\left(\frac{1}{\sqrt{|I|}}\sum_{i \in I} X_i > z  \right).
$$
The best response of the analyst to this naive testing rule involves selective reporting (``p-hacking''), where $I^* \in \argmax_{I\subseteq J} \act(X_I,I)$.
The problem with the naive test is that it does not control size.
Selective reporting by the analyst implies that the probability of rejection under the null is not bounded by $\alpha$.

We might correct for such selective reporting by making worst-case assumptions about all unreported statistics. 
This results in the \textbf{conservative test},
$$
    \act_3(X_I, I) = \bs 1\left(\frac{1}{\sqrt{2}}(X_1 + X_2) >  z \text{ and } I = \{1,2\}\right).
$$
If there are statistics that are not reported, then the null is not rejected.
This conservative test implies a probability of rejection given $X$ of 
$ P(J = \{1,2\}) \cdot \bs 1\left(\frac{1}{\sqrt{2}}(X_1 + X_2) > z\right).$
The conservative test controls size, but does not have good power properties.

As we show more generally in \autoref{sec:implementability} and \autoref{sec:testing} below, the \textbf{optimal test without a pre-analysis plan} can be implemented by selecting a full-data test of level $\alpha$.
When not all data are reported, the decision-maker needs to assume the worst about the unreported statistics, and then implements the corresponding full-data test.
The decision-maker can choose the full-data test to maximize (ex-ante) expected power, averaging over their prior for $\param$.

One possible full-data test ignores $X_2$, which is less likely to be observed in our numerical example, and rejects based on $X_1$ alone. This results in the test
$$
    \act_4(X_I, I) = \bs 1\left(X_1  >  z \text{ and } 1 \in I\right).
$$
This test implies a probability of rejection given $X$ of 
$ P(1 \in J)   \cdot \bs 1\left( X_1   > z\right).$
This test is optimal for some parameter values, while in general, the optimal test depends on the decision-maker's prior.\footnote{
    For the given prior over $J$, this test is for instance optimal when expected power is calculated using the degenerate prior $P(\param = .3)=1$. More generally, whether this rule is optimal depends on the prior for both $\param$ and $J$.
}
We lastly get to the \textbf{optimal test with a PAP}.
The optimal test with a PAP is of the same form as the optimal test without a PAP, except that the \textit{analyst} gets to choose the full data test, \textit{prior} to seeing any data.
Recall that in our example in this section the analyst knows the statistics $J$ that are available before possibly reporting a PAP, but we assume that they have no private information regarding $\param$ or $X$. (We relax these assumptions in our general setup below.)
The optimal implementable solution can be implemented as follows:
The analyst communicates which statistics are available by sending the pre-analysis message $M = J$, and the test is given by
$$
    \act_5(M, X_I, I) = \bs 1\left(\frac{1}{\sqrt{|M|}} \cdot  \sum_{i \in M} X_i > z   \text{ and } M\subseteq I\right).
$$
That is, the analyst commits to reporting all statistics in $J$, and for that set of statistics, the most powerful test is implemented.

\paragraph{Comparing size and power}

The left panel of \autoref{fig:power_curves} plots the power curves for the five testing rules, for $n= \dim(X) = 2$, which is the case that we have considered thus far. The right panel shows analogous plots for $n = 10$, where the probability $P(i \in J)$ of observing each of the statistics $X_i$ is evenly distributed over a grid from $.5$ to $.9$. The latter case illustrates the differences between testing rules more starkly.

A number of observations are worth emphasizing here.
First, the naive test does not control size. For $n=10$, the probability of rejection for $\param = 0$ is close to $.5$, instead of the nominal size of $.05$. This is due to selective reporting (``p-hacking'').
Second, the conservative test can be \textit{very} conservative.
Since it only rejects when all statistics of $X$ are reported, the probability of rejection under the alternative can be arbitrarily small, and remains below the nominal size of $.05$ for our example with $n=10$.
Third, the optimal test without a PAP does considerably better than either of these rules.
It controls size, and is in fact strictly conservative under the null.
At the same time, it has non-trivial power, which greatly exceeds that of the conservative test.
This test without a PAP remains itself far from optimal, however.
The optimal test with a PAP, lastly, controls size exactly, under the null. Furthermore, its power under the alternative considerably exceeds that of the optimal test without a PAP.

\paragraph{From our example to the general model}
Our motivating example is a special case of the general model that we lay out in \autoref{sec:model}.
The general model allows for cases where the researcher also has private information about $\param$, and where the researcher only has partial information about availability $J$ of the data. 
The general model also covers decision problems other than testing, including estimation and treatment choice.

%% file: Sections/Model.tex
\section{Setup}
\label{sec:model}

\begin{figure}[t]
  \centering
  \caption{Timeline}
  \label{fig:timeline}
  \bigskip
    \begin{tikzpicture}[node distance=3cm,
        every node/.style={fill=white, align=center, font=\footnotesize}]
      
      \draw[-{Latex[length=3mm]}] (0,0) -- (13,0); %
      
      \draw (2.5,-0.2) -- (2.5,0.2);
      \node[above] at (2.5,0.2) {Select $\M$ and \\ commit to $\act$};
      \draw (5,-0.2) -- (5,0.2);
      \node[below] at (5,-0.2) {Observe $\pi$, \\ send $M \in \M$};
      \draw (7.5,-0.2) -- (7.5,0.2);
      \node[below] at (7.5,-0.2) {Observe $(X_J,J)$, \\ select $I \subseteq J$};
      \draw (10,-0.2) -- (10,0.2);
      \node[above] at (10,0.2) {Observe $M$, $I$, $X_I$, \\ implement $A{=}\act(M,I,X_I)$};
      
      \node[align=left, text width=2cm] at (0,1) {Decision-maker};
      \node[align=left, text width=2cm] at (0,-1) {Analyst};
    \end{tikzpicture}
  \end{figure}
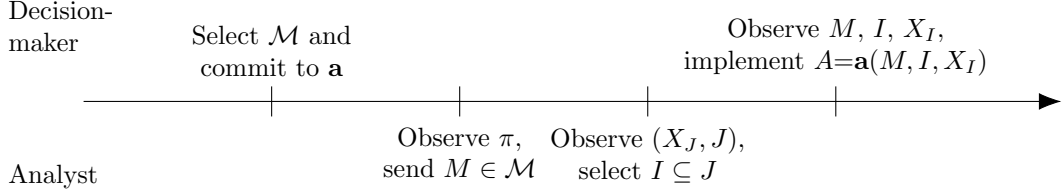

We next describe our general setup, which will be discussed for the rest of this paper.
Our setup consists of a game between a decision-maker and an analyst.
This game is summarized in \autoref{assumption:setup}.\footnote{Our notation does not distinguish explicitly between random variables and their realizations. This should not cause any ambiguity. Where the distinction is important, we point this out explicitly.}
The corresponding timeline is shown in \autoref{fig:timeline}.
Throughout, $X$ is a collection of statistics $X_i$, where $i \in \{1,\ldots,n\}$. $I$ and $J$ are (random) index sets, $I,J \subset \{1,\ldots,n\}$, and $X_I = (X_i)_{i\in I}$ denotes the subset of statistics corresponding to the index set $I$.
\begin{assumption}[Setup]
  \label{assumption:setup}
The game between decision-maker and analyst unfolds as follows:
\begin{enumerate}
  \itemsep0em
  \item The decision-maker selects a message space $\M$ and commits to a decision function $\act: (M,X_I,I) \mapsto A \in \mathcal{A}$.
  \item The analyst observes the private signal $\pi$ and sends a message $M \in \M$ to the decision-maker.
  \item The analyst observes the realization $(X_J,J)$ of available data and selects a subset $I \subseteq J$.
  \item The decision-maker observes the message $M$, the subset $I$, and the data $X_I$, and implements the decision $A = \act(M,X_I,I)$.
\end{enumerate}
The analyst and the decision-maker share a common prior $\P$ over the signal $\pi$, the parameter $\param$, the availability $J$, and the data $X$.
This prior satisfies that the conditional distribution of $X$ given $\param,J,\pi$ only depends on $\param$, i.e., $X|\param,J,\pi \stackrel{d}{=} X|\param$.
\end{assumption}

\paragraph{Discussion}
This is a game of partial verifiability.
The report $X_I$ is always truthful given $I$, but the non-availability of the statistics corresponding to $\{1,\ldots,k\} \setminus J$ cannot be verified by the decision-maker.
\emph{Selective reporting}, where not all available statistics are reported ($I \varsubsetneq J$), corresponds to p-hacking, or specification searching.
Mis-reporting of $X_I$, which corresponds to scientific fraud, is not allowed in our setting.

The private signal $\pi$ corresponds to \emph{analyst expertise}.
The signal $\pi$ might be informative about $\param$, corresponding to knowledge about which hypotheses are likely to be correct, about the likely magnitude of effect sizes, etc.
The signal $\pi$ might also be informative about $J$, corresponding to knowledge about the viability of different identification approaches, the availability of experimental sites, etc.

There is prior uncertainty of the decision-maker regarding the availability $J$ of statistics $X_i$.
Without such uncertainty, the mechanism design problem would be trivial, and the decision-maker could simply require the analyst to report everything, by threatening to take action $\min \mathcal A$ otherwise.
Prior uncertainty allows for \emph{``plausible deniability,''} because the decision-maker does not know the full set of results from which the reported results were selected.

In \autoref{assumption:setup}, we have left the message space $\M$ for the pre-analysis message $M$ unrestricted.
We will later encounter different, equivalent choices for $\M$:
The message $M$ might directly communicate the analyst signal $\pi$, or their corresponding posterior, in the spirit of the revelation principle in mechanism design.
Alternatively, and more realistically, the message $M$ might choose a decision function $\act$ from a restricted set, in the spirit of ``aligned delegation'' \citep{frankel2014aligned}. This latter formulation corresponds more directly to the practice of pre-analysis plans.

\paragraph{Objectives}
We have not yet described the objectives of either the decision-maker or the analyst; \autoref{assumption:setup} remains silent on these.
We allow for \emph{conflicting objectives}, which render the mechanism-design problem non-trivial.
By contrast, we have already imposed \emph{common priors}, so that there are no agency issues driven by divergent beliefs. 

We leave the decision-maker's objective unspecified at this point.
This allows us to first study implementability, as a general constraint on the set of decision-functions available to the decision-maker. 
This constraint does not depend on the decision-maker's objective.
We also do not impose that the decision-maker is an expected utility maximizer.
This allows us to study frequentist statistical decision problems subject to the constraint of implementability, including hypothesis testing %
and unbiased estimation, in addition to Bayesian decision problems.

By contrast, we do assume that the analyst is an expected utility maximizer.
We furthermore impose the following restriction on their utility function for most of our discussion.

\begin{assumption}[Monotonic analyst utility]
  \label{assumption:monotonicity}
  The decision $A$ is real-valued, $A \in \mathcal{A} \subset \mathbb R$.
  The analyst is an expected utility maximizer with utility $v(A)$, for a strictly monotonically increasing function $v$.
\end{assumption}

The analyst always prefers a higher outcome $A \in \mathcal{A}$.
In the context of testing, the analyst always prefers to reject the null hypothesis.
In the context of publication decisions, the analyst always would like their paper to be published.
In the context of drug approval, the pharmaceutical company always would like their drug to be approved.

%% file: Sections/Implementability.tex
\section{Implementability}
\label{sec:implementability}

Conventional statistical decision theory considers decision functions that map the available information into statistical decisions \citep{wald1950statistical,savage1951theory}.
In our context, such decision functions $\actreduced(\pi,X_J,J)$ map the signal $\pi$, the available data $X_J$, and the set $J$ of available statistics into decisions $A$.
We will call such functions $\actreduced$ \emph{reduced-form decision functions}.

In our setting, not all such decision functions are available to the decision-maker, because of analyst private information and conflicting objectives. 
In this section, we will characterize the set of \emph{implementable} reduced form decision functions $\actreduced$ which are consistent with analyst utility maximization.
This leads to constrained versions of conventional statistical decision problems, including hypothesis testing and point estimation.
We will show that implementation, in general, requires the use of pre-analysis messages.

\subsection{Which decision functions can be implemented?}
\label{subsec:implementable}

The analyst's optimal message $M^*$ and reported set  $I^*$ maximize analyst expected utility $\E[v(\act(M,X_I,I))]$,  given the decision rule $\act$.
Here $M^*$ and $I^*$ are random elements, where $M^*$ is measurable with respect to $\pi$, and $I^*$ is measurable with respect to $\pi,X_J,J$.
Analyst expected utility maximization and strict monotonicity of $v$ imply
\begin{align}
  I^* &\in \argmax_{I \subseteq J} \act(M^*,X_I,I), \text{ and}\nonumber
  \\
  M^* &\in \argmax_{M \in \mathcal{M}} \E[v(\act(M,X_{I^*},I^*))|\pi].
\end{align}

Consider now reduced-form decision functions $\actreduced(\pi,X_J,J)$ that map the information available to the analyst to a decision-maker action.
We say that a function $\actreduced$ is implementable if it is consistent with analyst utility maximization.

\begin{definition}[Implementable reduced-form decision rules]
  \label{def:implementability}
  A reduced form decision function $\actreduced(\pi,X_J,J)$ is \emph{implementable} if there exists a decision function $\act$ with best responses $M^*,I^*$ such that
  \[
    \actreduced(\pi,X_J,J) = \act(M^*,X_{I^*},I^*)
  \]
  almost surely.
\end{definition}

The following theorem provides a complete characterization of implementable reduced-form decision rules in our setting. The proof of this theorem, and all subsequent proofs, can be found in \autoref{sec:proofs}.\footnote{The function $\tilde \act$ is introduced in the following theorem as a technical device to deal with points $(\pi,X_J,J)$ outside the prior support.\\ It is worth noting that the revelation principle \citep{myerson1986multistage} does not directly apply to our setting, since misreporting of analyst ``types'' is constrained by the verifiability of their reports $(X_I,I)$, and by $I \subseteq  J$. See \cite{kephart2016revelation} for a discussion of the revelation principle under partial verifiability and, more generally, for settings where misreporting is potentially costly.}

\begin{theorem}[Implementability]
  \label{prop:implementability}
  Under Assumptions~\ref{assumption:setup} and \ref{assumption:monotonicity},
  a reduced-form decision function $\actreduced(\pi,X_J,J)$ is implementable if and only if there is some $\tilde{\act}$ such that $\actreduced(\pi,X_J,J) = \tilde{\act}(\pi,X_J,J)$ almost surely, and both of the following two conditions hold:
  \begin{enumerate}
    \item \textbf{Truthful message:} For all $\pi,\pi'$,
    \be
      \E[v(\tilde{\act}(\pi',X_J,J)) |\pi] \leq 
      \E[v(\tilde{\act}(\pi,X_J,J)) |\pi].
      \label{eq:properscoring}
    \ee
    \item \textbf{Monotonicity:} For all $\pi, X, J$ and $I \subseteq J$,
    \be
      \tilde{\act}(\pi,X_I,I) \leq \tilde{\act}(\pi,X_J,J).
      \label{eq:monotonicity}
    \ee
  \end{enumerate}
\end{theorem}

\autoref{prop:implementability} characterizes which reduced-form decision functions $\actreduced(\pi,X_J,J)$ can be implemented, but it does not tell us \emph{how} to implement them.
The following \autoref{prop:implementationalternatives} shows two different, canonical ways of implementing any such function.
The first implementation uses truthful revelation of analyst signals.
The second implementation uses delegation, where the analyst is allowed to choose the decision function from a pre-specified, restricted set $\B$. This second implementation corresponds closely to the actual practice of pre-analysis plans.
In this implementation, the analyst pre-specifies a mapping $b$ from the reported data $(X_J,J)$ to the decision $A = b(X_J,J)$.
\autoref{prop:implementationalternatives} shows that restricting attention to implementation by such pre-analysis plans is without loss of generality.

\begin{proposition}[Implementation]
  \label{prop:implementationalternatives}
  Under Assumptions~\ref{assumption:setup} and \ref{assumption:monotonicity}, a reduced-form decision rule $\actreduced$ can be implemented if and only if either of the following two conditions holds:
  \begin{enumerate}
    \item \textbf{Implementation by truthful revelation:}
      $\actreduced$ can be implemented with a decision rule $\act$ for which 
      $$\act(\pi,X_J,J) = \actreduced(\pi,X_J,J),$$
      where the message space is the set of all possible signals $\pi$.
    \item \textbf{Implementation by delegation (pre-analysis plan):}
    $\actreduced$ can be implemented with a decision rule $\act$ for which 
    $$\act(b,X_J,J) = b(X_J,J),$$
    where $b$ is restricted to lie in some set of functions $\B = \{b:\;(X_I,I) \mapsto \mathcal{A} \}$, chosen by the decision-maker, that acts as the message space.
  \end{enumerate}
\end{proposition}

\subsection{Alternative characterizations of implementability}

Having characterized implementable decision functions in general, we next discuss implementability for the special case of linear analyst utility $v$ and convex action space $\mathcal A$.
We then discuss the connection of truthful revelation to proper scoring.
We also consider variants of the model where decision-functions are constrained to be in some class of suitably simple functions.

\paragraph{The set of implementable rules as a convex polytope}
In addition to Assumptions~\ref{assumption:setup} and \ref{assumption:monotonicity}, assume for a moment that the action space $\mathcal{A} \subseteq \R$ is convex, and that analyst utility is linear -- without additional loss of generality, $v(A) = A$.
The leading examples involve binary decisions, where we interpret $A$ as the \emph{probability} of a positive decision. 
Binary decisions occur for statistical testing, as discussed in \autoref{sec:testing} below, as well as for publication decisions, drug approval, etc.
Linearity is without loss of generality for the case of binary decisions; in this case, it follows from expected utility maximization. 
Suppose finally that $\pi$ has finite support.

Under these additional assumptions, we get that every implementable decision functions $\actreduced$ is almost surely identical to a function $\tilde{\act}$ in the convex polytope characterized by the following constraints:
\bals
  \tilde{\act}(\pi,X_J,J) &\in \mathcal{A},&&&\text{(Support)}\\
  \tilde{\act}(\pi,X_{I},{I})- \tilde{\act}(\pi,X_{J},{J}) &\leq 0 &\forall\; \pi, X_J, J, I\subseteq J, &&\text{(Monotonicity)}\\
  \sum_{X_J,J} \left(\tilde{\act}({\pi'},X_J,J){-}\tilde{\act}({\pi},X_J,J)\right)\: \P_{\pi}(X_J,J) &\leq 0 &\forall\; \pi', \pi.&&\text{(Truthful message)}
\eals
In the last inequality, $\P_\pi$ is a shorthand for the analyst's posterior distribution conditional on $\pi$.
This characterization of the implementable set follows immediately from \autoref{prop:implementability}.
    
If, furthermore, the decision-maker objective is linear in $\actreduced$, as is the case for a Bayesian decision-maker and binary actions, or if it is linear with an additional linear constraint, as is the case for expected power maximization subject to size control, then the problem of finding the optimal implementable reduced form decision function becomes a linear programming problem.
Efficient algorithms exist for numerically solving such problems, cf. \cite{vanderbei2020linear}.
We will return to this point in \autoref{sec:testing} below.
We leverage such linear programming algorithms in our interactive app for finding optimal PAPs.
    
\paragraph{Truthful revelation of beliefs and proper scoring}
Condition~\eqref{eq:properscoring} in \autoref{prop:implementability} ensures that the analyst reveals their relevant prior information truthfully.
Condition~\eqref{eq:properscoring} is equivalent to the definition of a proper scoring rule, as introduced by \cite{savage1971elicitation}.
The theory of proper scoring rules has regained importance in the more recent statistics and machine learning literature, cf. \cite{gneiting2007strictly}.

Let us elaborate on this equivalence.
Given a reduced form decision rule $\actreduced$, define
\be
  S(\pi', \pi) = \E_\pi[v(\actreduced(\pi',X_J,J))].
  \label{eq:definitionscore}
\ee
The expectation $\E_\pi$ is taken over the conditional distribution $\P_\pi$ of $X_J,J$ given $\pi$.
Here we assume for simplicity that $X$ has finite support, though the argument generalizes.
Denote the Euclidean inner product for functions of $X_J,J$ by
$\langle f( \cdot ), g( \cdot ) \rangle = \sum_{X_J,J} f(X_J,J) \cdot g(X_J,J)$, where the running indices $X_J,J$ are understood here as values, rather than random variables.
$P_\pi$, the distribution of $(X_J,J)$ given $\pi$, is a vector in the space on which this inner product is defined.
We obtain the following characterization, which was first stated by \cite{savage1971elicitation} and is restated as Theorem 2 in \cite{gneiting2007strictly}.

\begin{proposition}[Proper scoring rule]
  \label{prop:proper}
  Condition~\eqref{eq:properscoring}, the truthful message condition, holds for all $\pi,\pi'$ if and only if there exists a convex function $G$ of $P_\pi$, with sub-gradient $G'$, such that $G(\P_\pi) = S(\pi,\pi)$ on the support of $\pi$, and such that
  \(
    S(\pi', \pi)
      = G(\P_{\pi'}) + \langle G'(\P_{\pi'}, \cdot),  \P_\pi - \P_{\pi'}\rangle .
  \)
\end{proposition}

\paragraph{Simple pre-analysis plans}
Item~2 of \autoref{prop:implementationalternatives} shows that reduced form decision rules can be implemented by delegation:
The decision-maker offers a set  $\B = \{b:\;(X_I,I) \mapsto \mathcal{A} \}$ of permissible pre-analysis plans (decision functions).
The analyst then chooses and communicates one of the decision functions $b \in \B$ before gaining access to the data.

In practice, some pre-analysis plans may be unrealistically complicated, and we may wish to restrict attention to a smaller set $\B_0$ of simpler mappings.
The decision-maker could be restricted to choosing $\B$ as a subset of this set of simple mappings, $\B \subseteq \B_0$.

One example of such a restricted set $\B_0$ are the index rules implemented in our app, which is described below.
These index rules are of the form $$b(X_I,I) = \bs 1\left(I_b\subseteq I \text{ and } \sum_{i\in I_b} X_i \geq z_b\right),$$ where $I_b$ is the set of statistics included in the index, and $z_b$ is a critical value.

\subsection{Are pre-analysis messages needed?}

\paragraph{Aligned objectives}
Why does implementability in our setting require a pre-analysis message, if that is not the case in conventional statistical decision theory?
Assume for a moment that analyst and decision-maker share the same objective function.
In this case, is there any need for a \emph{pre}-analysis message?
The answer is no.

To see this, consider the following variant of our setup.
Suppose everything is as in \autoref{assumption:setup} (\autoref{fig:timeline}), except that the analyst gets to choose the message $M$ \emph{after} they observe the data $X_J,J$.
Put differently, the analyst cannot provide a verifiable time-stamp for their message $M$ to the decision-maker.
The following observation states that in this modified setting, where there is no \emph{pre}-analysis message,  the decision-maker can still implement the first-best reduced-form decision rule, provided that preferences are aligned.

\begin{proposition}[First-best decisions for aligned preferences]
  \label{prop:firstbest}
  Under the modified \autoref{assumption:setup} where the message $M$  can depend on the realization of $(X_J,J)$, assume that analyst and decision-maker are expected utility maximizers who share the same utility function $u(\Act,\param)$.
  Then the decision-maker's first-best reduced-form decision rule $\actreduced(\pi,X_J,J)$ is implementable. 
\end{proposition}

As \autoref{prop:firstbest} shows, \emph{pre}-analysis messages only become potentially useful in the presence of both private information \emph{and} misaligned preferences.

\paragraph{Implementability without pre-analysis message}

We next characterize the set of decision functions $\actreduced$ that are implementable without a pre-analysis message, when objectives can be misaligned.
In this case, the implementable functions are exactly the functions $\actreduced(\pi,X_J,J)$ that satisfy monotonicity, with respect to set inclusion for the index set $J$, and that do not depend on $\pi$.
Analyst expertise can thus not be used to improve decisions \emph{at all}, in the absence of a pre-analysis message.
The proof of the following proposition parallels the proof of \autoref{prop:implementability}.

\begin{proposition}[Implementability without pre-analysis message]
  \label{prop:monotonicity}
  Under Assumptions~\ref{assumption:setup} and \ref{assumption:monotonicity}, with the additional
 constraint that there is no pre-analysis message,
  a reduced-form decision function $\actreduced$ is implementable if and only if 
  there is a function $\tilde{\act}$ with almost surely $\actreduced(\pi,X_J,J) = \tilde{\act}(X_J,J)$ and
  \begin{align}
    \label{eq:setmonotonicity}    
    \tilde{\act}(X_I,I) \leq \tilde{\act}(X_J,J)
  \end{align}
  for almost all $X,J$ and all $I \subseteq J$.
\end{proposition}

%% file: Sections/Testing.tex
\section{Frequentist hypothesis testing}
\label{sec:testing}

We next specialize our general framework to the setting of frequentist hypothesis testing.
In this setting, the decision-maker decides whether to reject a null hypothesis.
We assume that the decision-maker wants to maximize expected power subject to size control.
The analyst, however, always prefers a rejection of the null hypothesis.

Building on our previous results, we characterize the set of implementable testing rules that satisfy size control, in \autoref{subsec:testpaps}. We furthermore provide a simple mechanism that allows the decision-maker to implement the optimal testing rule. This mechanism requires a pre-analysis plan, where the analyst may choose any full-data test that satisfies size control, and the decision-maker makes worst-case assumptions about any unreported data.
This mechanism solves the decision-maker's problem.

In \autoref{subsec:discretetests} we then consider the analyst's problem of finding an optimal response to this mechanism, and show that they have to solve a linear programming problem to find the optimal pre-analysis plan. We provide software to solve this problem of the analyst.
We also characterize the set of possible solutions to the analyst's problem, by describing the set of extremal points of their feasible set.

Throughout, we focus on the problem of testing a single (joint) hypothesis, and leave an extension to deciding which of multiple hypotheses to reject for future work.

\subsection{Decision-maker and analyst objectives}
Assume that the decision $A \in [0,1]$ represents the probability, given $(M,X_J,J)$, of rejecting the null hypothesis $\param \in \Theta_0$. 
Suppose that the analyst is an expected utility maximizer, who (ex-post) only cares about the binary testing decision. 
Ex-ante, the analyst thus wants to maximize expected power.
It follows that their utility is linear in $A$. We can then make the following normalizing assumption, without loss of generality.
\begin{assumption}[Power analyst utility]
    \label{assumption:power}
    Analyst utility is
    \[
      v(A) = A.
    \]
\end{assumption}
The decision-maker also wants to maximize expected power, but subject to the constraint of size control under the null hypothesis.

\begin{definition}[Size control]
  \label{def:size}
  We say that a reduced-form decision rule $\actreduced$ which satisfies $0 \leq \actreduced \leq 1$ controls size at level $\alpha \in (0,1)$ if
  \begin{align}
    \label{eq:sizecontrol}
    \sup_{\mathclap{\pi,\param \in \Theta_0, J \subseteq \{1,\ldots,n\}}}   \E[\actreduced(\pi,X_J,J) | \param,\pi,J] &\leq \alpha.
  \end{align}
\end{definition}

Recall that we imposed, in \autoref{assumption:setup}, that the conditional distribution of $X$ only depends on $\param$, that is, $X|\param,J,\pi \stackrel{d}{=} X|\param.$
Under this assumption, the conditional expectation $\E[\actreduced(\pi,X_J,J) | \param,\pi,J]$ is well-defined even outside the joint support of $\pi,\param,J$, as long as $\param$ is within its marginal support.

\subsection{Decision-maker solution: Pre-specified full-data tests}
\label{subsec:testpaps}

The implementability results of \autoref{sec:implementability} allow us to characterize optimal pre-analysis plans for hypothesis testing as follows.
\begin{theorem}[Optimal pre-analysis plans with size control]
  \label{prop:delegation}
  Define $\mathcal T$ to be the class of measurable full-data tests $t: \X \rightarrow [0,1]$ satisfying size control, $\sup_{\param \in \Theta_0} \E[t(X) | \param] \leq \alpha$.
  Under \autoref{assumption:setup}, \autoref{assumption:monotonicity}, and \autoref{assumption:power},
  the power-maximizing decision rule subject to the constraints of implementability (\autoref{def:implementability}) and size control (\autoref{def:size}) can be implemented by requiring the analyst to communicate, as a pre-analysis message, a full-data test $t \in \mathcal T$, 
  and then rejecting the null with conditional probability
  \[b(X_I,I) = \inf_{X';\;X'_I=X_I} t(X').\]
\end{theorem}

This result builds on the general characterizations of \autoref{prop:implementability} and \autoref{prop:implementationalternatives}.
To get further intuition for \autoref{prop:delegation} note, first, that it is sufficient to verify size control for the \textit{full-data} test $t$.
The reason is that implementable reduced-form decision rules must fulfill the monotonicity constraint \eqref{eq:monotonicity}.
Subject to monotonicity in $I$, size control of $\actreduced$ in the sense of \autoref{def:size} is equivalent to size control for the full-data test $\actreduced(\pi, X, \{1,\ldots,k\})$.

Note, second, that for \emph{optimal} reduced-form testing rules the monotonicity constraint is in general binding, since both decision-maker and analyst aim to maximize expected power, subject to the constraints.
For optimal rules it is therefore without loss of generality to assume  $\actreduced(\pi,X_J, J) = \inf_{X';\;X'_J=X_J} t(X')$, which can be implemented by $b$ as in the statement of the theorem.

\subsection{Analyst solution: Linear programming}
\label{subsec:discretetests}

\autoref{prop:delegation} solves the optimal testing problem from the decision-maker's perspective: Let the analyst pre-specify a valid full-data test, and then make worst-case assumptions about unreported data.
We next turn to the analyst's problem: What full-data test should they specify?
This problem can be cast as a linear programming problem.
The optimal value for any linear programming problem can be achieved on the set of extremal points of the feasible set.%
\footnote{The same holds more generally, for the maximum of a convex function on a convex set.}
This insight, which is of central importance to mechanism design \citep{mechanismdesignnotes2023}, allows us to characterize the set of potential solutions to the optimal testing problem subject to implementability.

\paragraph{Linear objective and linear feasible set}

For ease of exposition, we focus on point null hypotheses \(\Theta_0 = \{\param_0\}\) in the following.
Our results extend to compound hypotheses. %
Denote $K=\{1,\ldots,k\}$ the index set of all potentially available statistics.
Let $\mathcal B$ be the set of measurable functions $b(X_J,J)$ defined by the following constraints.
\bal
\int  b(X,K) d\P_{\param_0}(X) &\leq \alpha, &&&\text{(Size control)}\nonumber \\
b(X_J,J)&\in [0,1] & \forall\; J,X,&&\text{(Support)}\nonumber \\
b(X_J,J) &\leq b(X,K) & \forall\; J,X.&&\text{(Monotonicity)}
\label{eq:mathcalB}
\eal
This is the set of testing rules from which the analyst is effectively allowed to choose, after observing their private signal $\pi$.
This characterization applies to both discrete and continuously distributed $X$.
The set $\mathcal B$ is a convex polytope.

The (interim) analyst objective function is given by expected power, conditional on their private signal $\pi$,
\bals
  \E_\pi[b(X_J,J)] &= \int b(X_J,J) d\P_\pi(X,J).&&\text{(Interim expected power)}
\eals
We provide code, in the form of an interactive app, which allows the analyst to easily solve the problem of maximizing expected power, subject to $b\in \B$.\footnote{This app is available at \href{https://maxkasy.github.io/home/pap_app/}{https://maxkasy.github.io/home/pap\_app}.
}

\paragraph{The case of known $J$}
The analyst's problem simplifies to the standard problem of finding a test of maximal expected power subject to size control, if we assume that the analyst knows the value of $J$, at the time of specifying their PAP. 
Let $J'$ be this known non-random value of $J$.
Under this assumption, the optimal implementable test is a function of $X_{J'}$ only, and can be written as a likelihood ratio test.
\begin{proposition}
  \label{prop:knownJ}
  Suppose that \autoref{assumption:setup}, \autoref{assumption:monotonicity}, and \autoref{assumption:power} hold, and consider the mechanism specified in \autoref{prop:delegation}.
  Suppose additionally that $P_\pi(J=J') = 1$ for some non-random value $J'$. 
  Then there exists a solution $b$ to the analyst's problem such that $b(X_K, K) = b(X_{J'}, J')$ for all values of $X$. 
  
  Any solution of the analyst's problem that is of this form furthermore satisfies that 
  $$b(X_K, K) = \begin{cases}
    1 & \text{when } d\P_{\pi}(X_{J'}, J') > \kappa  \cdot d\P_{\param_0}(X_{J'}, J') \\
    0 & \text{when } d\P_{\pi}(X_{J'}, J') < \kappa  \cdot d\P_{\param_0}(X_{J'}, J')
  \end{cases}.
  $$  
  for some critical value $\kappa$.
\end{proposition}
\autoref{prop:knownJ} implies that the null should be rejected based on the value of the likelihood ratio test statistic $\frac{d\P_{\pi}(X_{J'}, J')}{d\P_{\param_0}(X_{J'}, J')}$ (assuming this statistic is well defined). Note that the likelihood in the numerator $d\P_{\pi}(X_{J'}, J')$ is in fact the \textit{marginal} likelihood under the interim prior given $\pi$, averaging over both the interim prior for $\param$, $J'$, and over the sampling distribution of $X$ given $\param$.
See \cite{lehmann2006testing} (Section 3.8) for a discussion of statistical tests that maximize weighted average power.

\paragraph{Potentially optimal tests: Extremal points of $\mathcal B$}
Let us now return to the more general case, where the analyst does not necessarily know the value of $J$ after observing $\pi$.
Suppose we maintain \autoref{assumption:setup}, \autoref{assumption:monotonicity}, and \autoref{assumption:power}, but impose no further assumptions on the (interim) prior $\P_\pi$ of the analyst. %
What can we say about the set of potential solutions $b$ to the analyst's problem, in this case?
The following proposition provides a characterization, based on the set of extremal points of the set  $\mathcal B$, intersected with the set of rules $b$ for which monotonicity is binding.

\begin{proposition}$\;$
  \label{prop:extremal}
  \begin{itemize}
    \item Suppose that \autoref{assumption:setup}, \autoref{assumption:monotonicity}, and \autoref{assumption:power} hold, and consider the mechanism specified in \autoref{prop:delegation}.
    Then there exists a full-data test $t$  which is a best response of the analyst such that
    $b(X_J, J) = \inf_{X':\; X'_J=X_J} t(X')$
   is extremal in $\B$.
   
   \item Suppose additionally that $t$ takes on a finite number of values. Then a function $b$ of this form is extremal in $\mathcal B$ if and only if the following conditions hold:
   \begin{enumerate}
    \item $t(X) \in \{0,q,1\}$ for all $X$, for some $0<q<1$.
    \item If there exists  $X$ such that $t(X) = q$, then $\P_{\param_0}(t(X) = q) > 0$.
    \item For any $X \neq X'$ such that $t(X) = t(X') = q$, there exists a value $J$ such that $X_J = X'_J$ and $b(X_J,J)= b(X'_J, J) = q$.
   \end{enumerate}
  \end{itemize}
\end{proposition}

In other words, we can restrict our attention to testing rules that partition values of the data $X$ into at most three regions: one where the test always rejects; one where the test never rejects; and one where it rejects with a single, intermediate probability.
Furthermore, if there is more than one value for which the test takes this intermediate rejection probability,
then the monotonicity constraint in the construction of the tests $b$ is binding for at least some subset $J$.

The result in \autoref{prop:extremal} characterizes the set of extremal points of $\mathcal B$ for which monotonicity is binding. The optimal analyst response is necessarily in this set.
Can all of these points be rationalized as optimal for some analyst interim prior?
The following proposition provides a partial answer.

\begin{proposition}
  \label{prop:rationalizing}
  Consider $b \in \mathcal B$ with $\P_{\param_0}(b(X,K) \notin \{0,1\}) = 0$, and such that the size constraint is binding.
  Then there exists a prior $\P_\pi(X_J,J)$ such that $b$ maximizes the objective
  $
    \int b(X_J,J) d\P_\pi(X_J,J)
  $
  in $\mathcal B$.
\end{proposition}
This result shows that all testing rules that control size without an intermediate probability of rejection can be rationalized.

%% file: Sections/Case_studies.tex
\section{Case study}
\label{sec:casestudies}

We next discuss a numerical example, to illustrate our results on optimal pre-analysis plans for hypothesis testing.
Our example is calibrated to the data and the priors reported in \cite{dellavigna2018motivates}, who experimentally evaluate 15 different treatments to induce costly effort, in addition to 3 control treatments.
The outcome $X_i$ is the effect of treatment $i$ on the average number of button presses, in an Amazon Mechanical Turk task.
We consider the effect relative to the control treatment where participants are paid 1 cent per 100 button presses.
\cite{dellavigna2018motivates} also report prior predicted treatment effects, as elicited from 208 academic experts.

We use these expert predictions to calibrate our prior for $\param = (\param_i)_{1\leq i \leq 15}$, where $\param_i$ is the true effect of treatment $i$.
We assume that $\param \sim N(\mu, \Sigma)$ is jointly normal, with prior mean $\mu$ equal to the average of expert forecasts, and prior variance $\Sigma$ equal to the variance across forecasts.
We furthermore assume that the estimated treatment effects have a sampling distribution of $X_i \sim \N(\theta_i, \sigma^2_i / n + \sigma_0^2 / n)$, where the sample size is $n = 100$, and $\sigma_0^2 / n$ is the variance of the mean outcome for the control treatment.\footnote{The variation across experts is different from the variance of the prior of any individual expert. We furthermore deviate the original sample size of around 550 per arm in the paper. For both these reasons, or numerical example should only be thought of as a calibration for the purpose of illustrating our theory.} The standard deviations $\sigma_i$ are assumed to be known and correspond to the standard errors reported in \cite{dellavigna2018motivates}.

We lastly assume, for the purpose of illustration, that the analyst only intends to run experiments for two of the 15 experimental treatments, corresponding to arm 1 (4 cents per 100 presses), and arm 2 (a lottery with a chance of winning 1 dollar per 100 presses with 1\% probability).
We assume, for now, that the analyst knows ex-ante that $J=\{1,2\}$.
Consider the null (joint) null hypothesis that there are no treatment effects for any of the incentive schemes, $\param_i = 0$ for all $i$.

\begin{figure}[p] 
    \centering 
    \caption{Analyst knows $J$} 
    \begin{subfigure}[b]{0.49\linewidth} \centering \includegraphics[width=\linewidth]{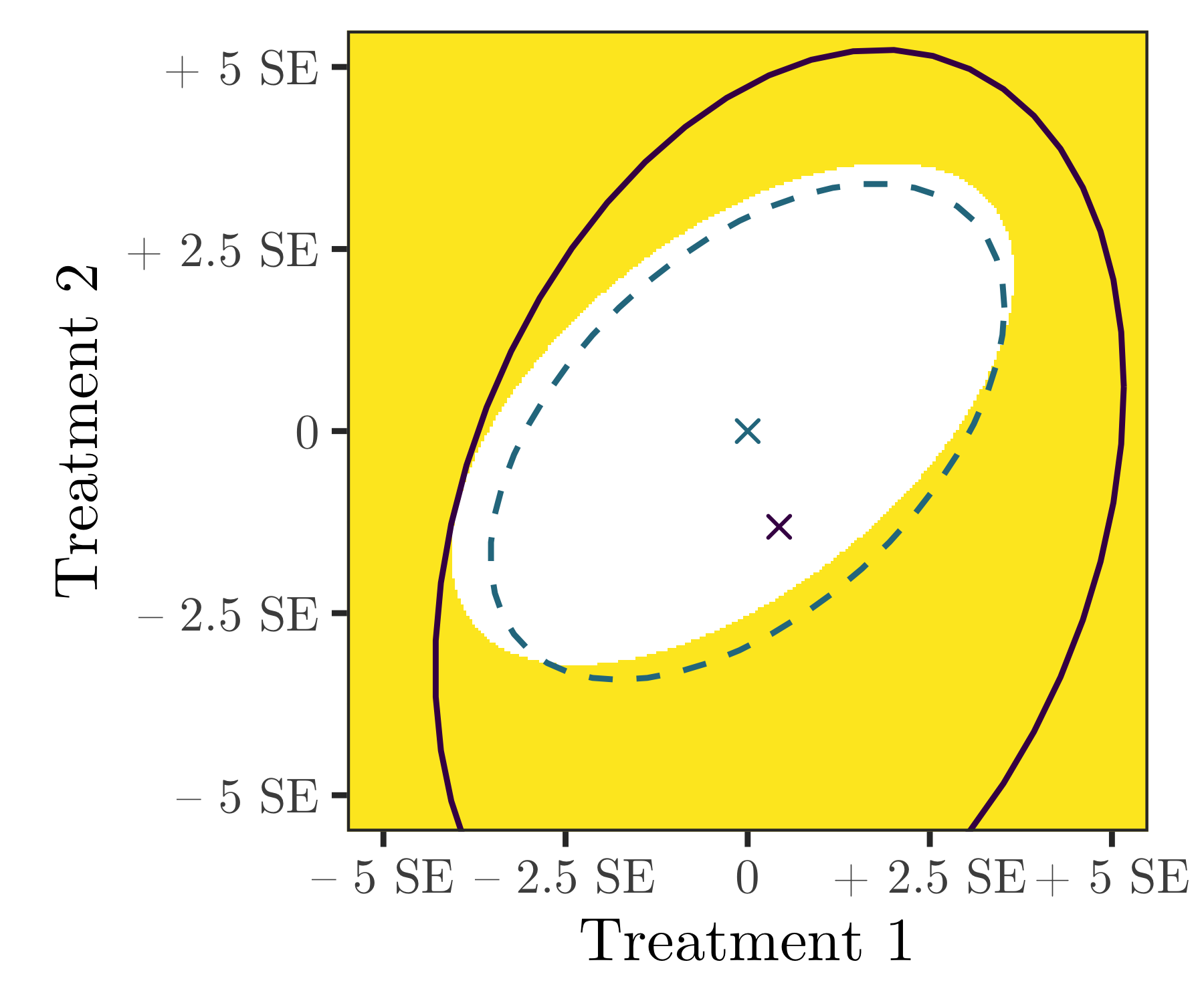} 
        \caption{Optimal PAP} 
        \label{fig:illustration0_complex} 
    \end{subfigure} 
    \begin{subfigure}[b]{0.49\linewidth} 
        \centering 
        \includegraphics[width=\linewidth]{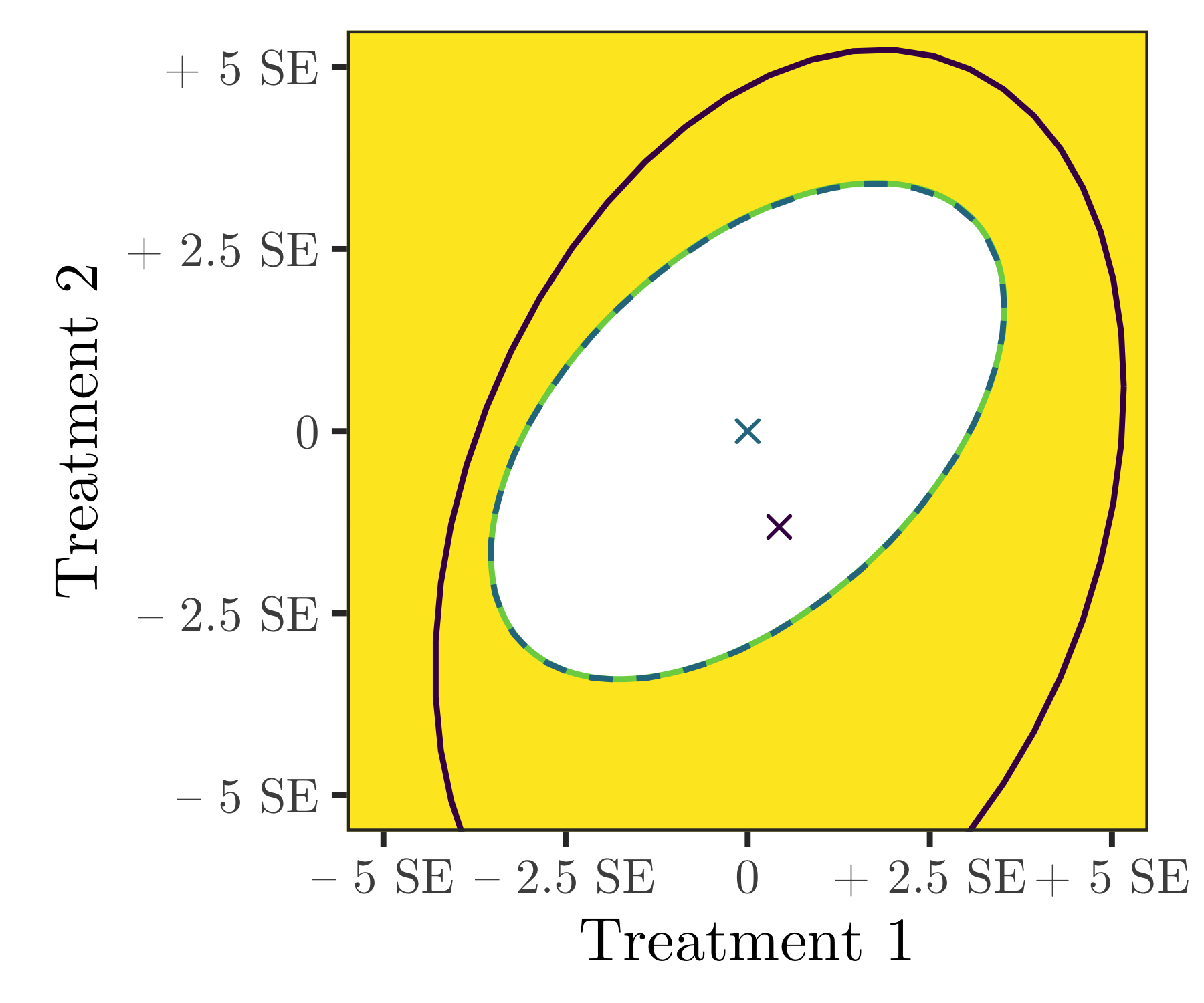} 
        \caption{Optimal simple PAP} 
        \label{fig:illustration0_trivial} 
    \end{subfigure} 
    
    \bigskip
    
    \caption{Analyst is uncertain about $J$} 
    \begin{subfigure}[b]{0.49\linewidth} \centering \includegraphics[width=\linewidth]{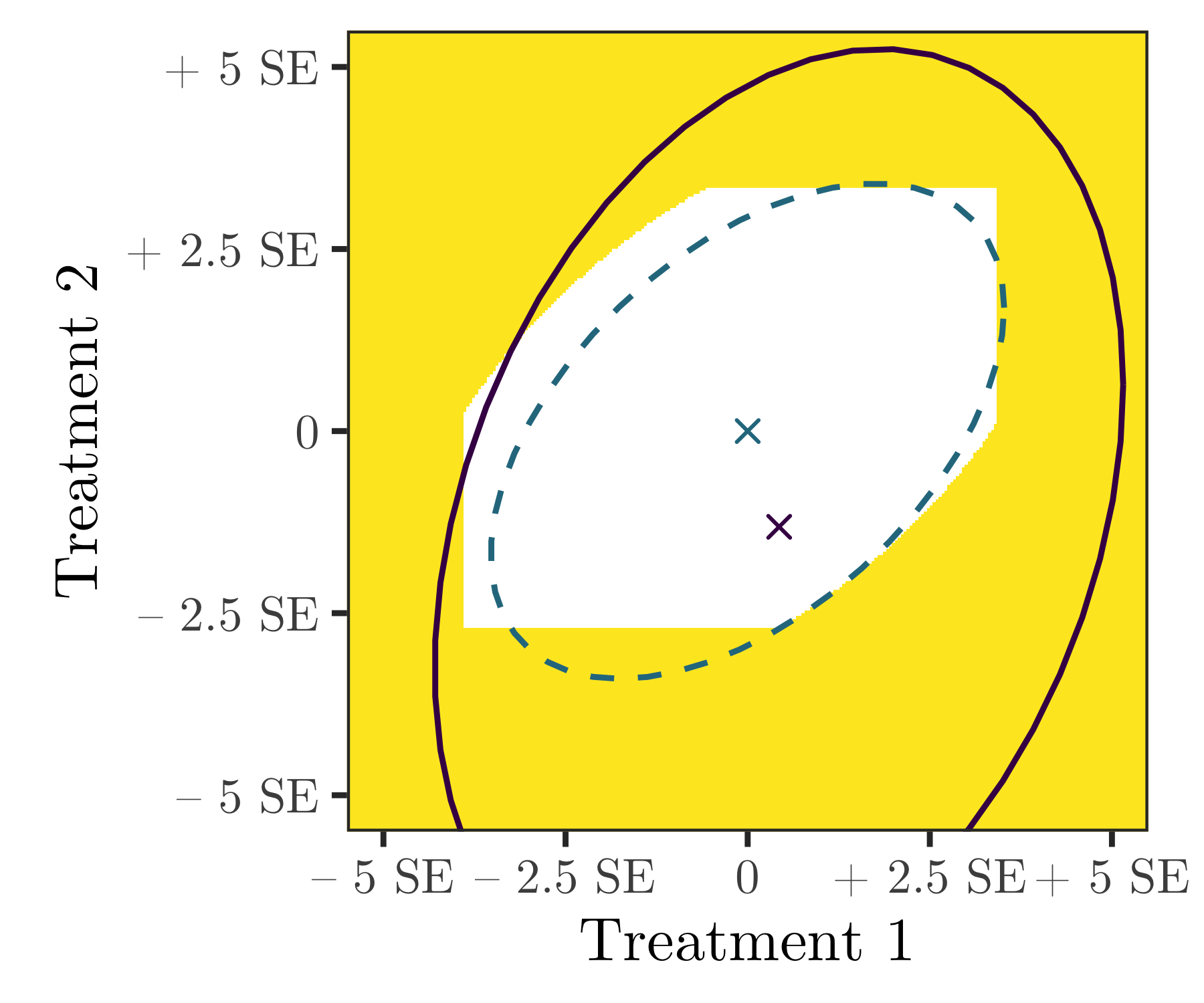} 
        \caption{Optimal PAP} 
        \label{fig:illustration1_complex} 
    \end{subfigure} 
    \begin{subfigure}[b]{0.49\linewidth} 
        \centering 
        \includegraphics[width=\linewidth]{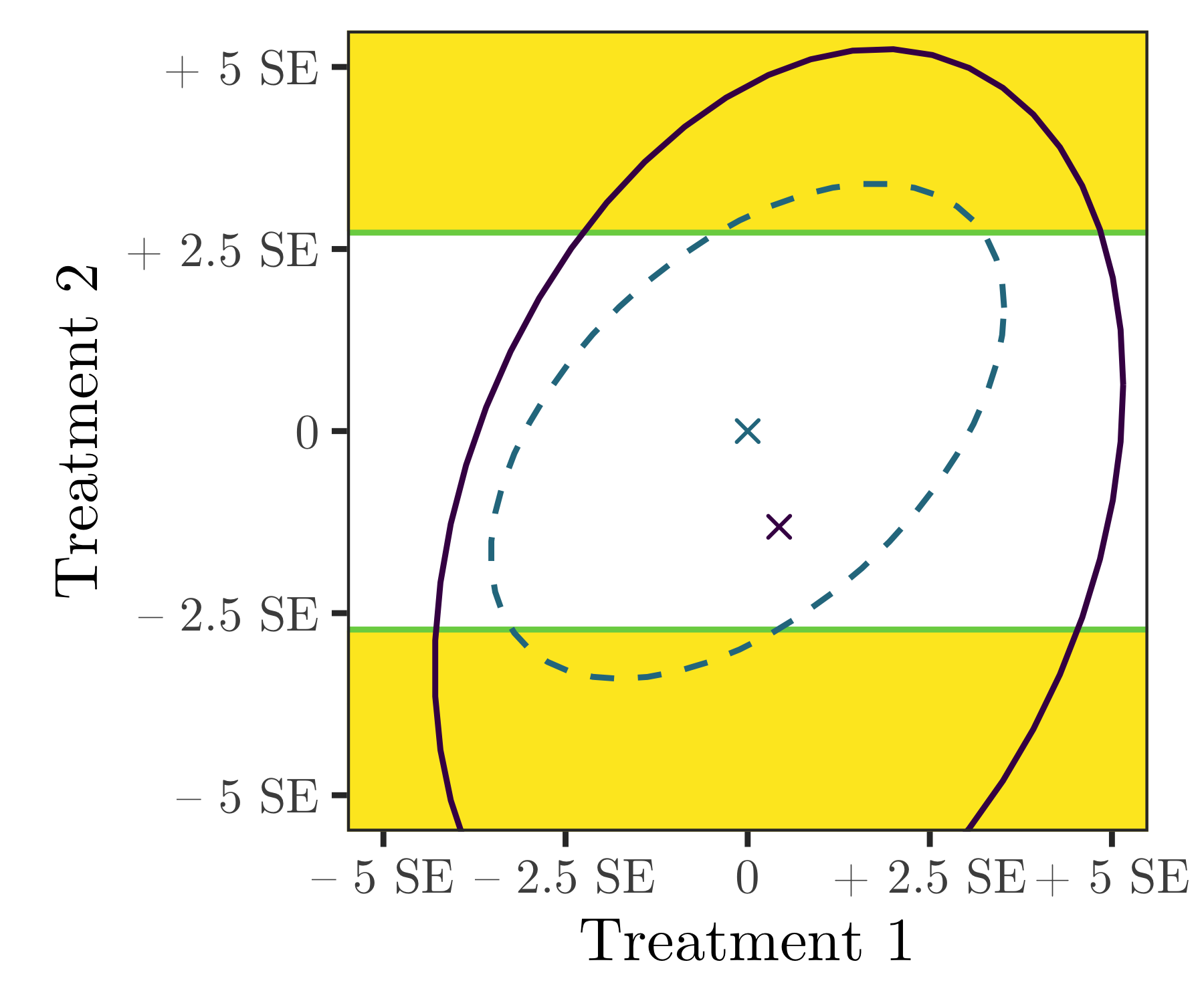} 
        \caption{Optimal simple PAP} 
        \label{fig:illustration1_trivial} 
    \end{subfigure} 

\end{figure}

\paragraph{The optimal PAP}
What is the optimal PAP for this null?
The answer is given by \autoref{prop:knownJ}.
The optimal PAP, for known $J$, pre-specifies a test which rejects whenever both components of $J$ are reported, and $\log\left(\tfrac{d\P_{\pi}(X_{J}, J)}{d\P_{\param_0}(X_{J}, J)}\right)$ exceeds some critical value.
Under our assumptions,
\bals
    \log\left(\tfrac{d\P_{\pi}(X_{J}, J)}{d\P_{\param_0}(X_{J}, J)}\right) &=
    const. + \psm{X_1 - \mu_1 \\ X_{2} - \mu_{2}}' S^{-1} \psm{X_1 - \mu_1 \\ X_{2} - \mu_{2}} - 
             \psm{X_1 \\ X_{2}}' S_0^{-1} \psm{X_1 \\ X_{2}} %
\eals
where $S_0$ is the sampling variance of $X_J$, and $S$ is the prior variance of $X_J$, which equals the sum of the prior variance of $\param_J$ plus the sampling variance,
\bals
S_0 &= \tfrac1n \psm{\sigma_0^2 + \sigma_1^2 & \sigma_0^2 \\ \sigma_0^2 & \sigma_0^2 + \sigma_{2}^2}, &
S = \psm{ \Sigma_{1,1} & \Sigma_{1,2} \\ \Sigma_{2,1} & \Sigma_{2,2}} + S_0.
\eals
We visualize this test in \autoref{fig:illustration0_complex}.
The axes of the graph represent estimated treatment effects, normalized by their sampling standard error, $X_i / \sqrt{(\sigma_0^2 + \sigma_i^2)/n}$, for $i=1,2$.
The blue ellipse (dashed line) represents the null distribution of $X_J$, with 95\% of draws falling within the circle; and the purple ellipse (solid line) represents the prior marginal distribution of $X_J$.
The optimal rejection region at a 5\% size is shaded in yellow. The likelihood-ratio test of \autoref{prop:knownJ} yields an ellipsoidal rejection region.

\paragraph{An optimal simple PAP}
In practice, fully optimal tests may be hard to describe in a PAP. 
What is the optimal PAP subject to an additional simplicity constraint?
Let us restrict attention to tests that reject if the test statistic
$X_J'  \cdot S_{0}^{-1} \cdot  X_J$ exceeds some critical value $c_J$ and if all components in $J$ are reported, where both $J$ and the critical value are pre-specified.
This is the standard Wald ($\chi^2$) test for the subset $J$.
Subject to this restriction, it is optimal for the analyst to pre-register their true $J = \{1,2\}$, and a critical value of $6$ (for a test of size .05).
We visualize this test in \autoref{fig:illustration0_trivial}.
Restricting tests to be simple leads to a loss in average power, but this loss is small in our numerical example. Average power of the optimal test is approximately $.52$.
The restriction to a Wald test reduces power to $.50$.

\paragraph{Analyst uncertainty about $J$}
Assume now that the analyst is uncertain about which components $J$ will be available.
Maybe some experiments are not always feasible, or data collected differ from those in the original plan.
Assume that arm 1 is available with ex-ante probability .5, and arm 2 with probability .7, independently across arms.
In this case, the rejection regions of the optimal PAP are more complex, and are given by the solution to the linear programming problem discussed in \autoref{subsec:discretetests}.
\autoref{fig:illustration1_complex} plots the optimal test, which solves this linear program.
If one of the arms is not available in the end, then the decision-maker makes worst-case assumptions about this arm, and implements the corresponding testing decision.
Because the components $i$ are not always available, overall expected power only equals $.32$ in this example.

\begin{figure}[t]
    \centering
    \caption{Which treatment arms should be registered?}
    \includegraphics[width=0.7\linewidth]{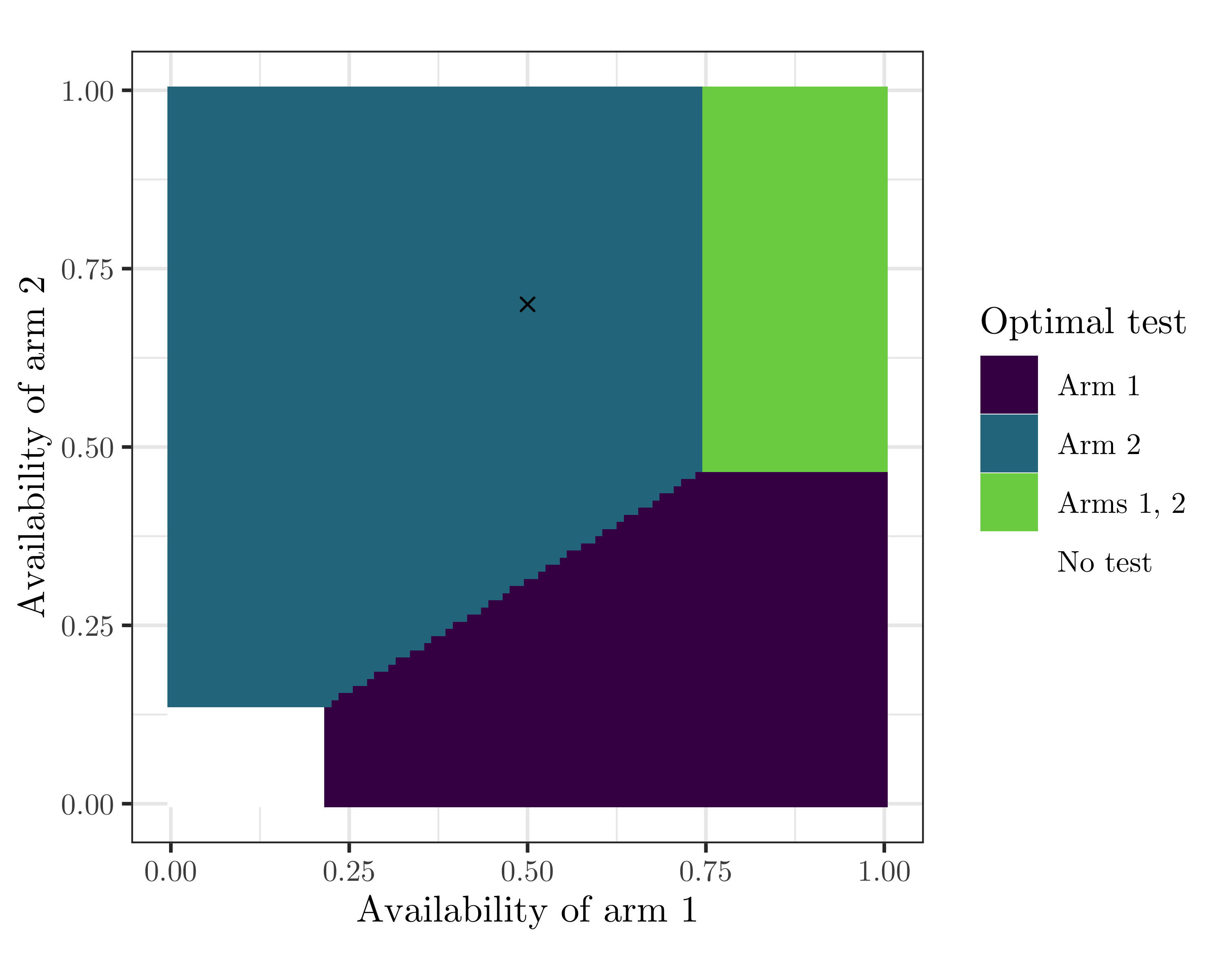}    
    \label{fig:illustration_bestest}
\end{figure}
We can also, again, consider simple PAPs, which specify Wald tests for some pre-selected set of components $J'$.
The optimal simple PAP ignores arm 1 and specifies a two-sided t-test that rejects for $\frac{\sqrt{n} \left| X_{2}\right|}{\sqrt{\sigma^2_{2} + \sigma^2_0}} > 1.96$ (\autoref{fig:illustration1_trivial}).
That is, despite arm 1 being available some of the time, it is better to only consider arm 2 in this case.
This result is driven by the different priors over the effect of these treatments, as well as by different availabilities, where arm 2 is more likely to lead to a rejection and is more likely to be available.
Restricting attention to such a simple test reduces expected power from .32 (for the optimal test) to only .26.

Concluding our discussion of this numerical example, we plot  in \autoref{fig:illustration_bestest} how the optimal set of pre-registered components $J'$, for a simple test, depends on the probability that data for either treatment is available.
\autoref{sec:casestudy_simple} elaborates further.

%% file: Sections/Conclusion.tex
\section{Conclusion}
\label{sec:conclusion}

We conclude by summarizing our main contributions, before discussing some limitations of our model and avenues for future research.
We have proposed a principal-agent model of pre-specification in empirical research.
In our model, a decision-maker relies on the examination and reporting of data by an analyst.
The analyst can selectively report statistics that they observe, but they cannot lie about the observed statistics.
The decision-maker does not know which data are available to the analyst. This allows for plausible deniability.

Our model provides a theoretical justification for pre-analysis plans (or, more generally, pre-analysis messages), which cannot be rationalized in traditional single-agent statistical decision theory. 
There is no need for sending messages prior to seeing data in the single agent framework -- in fact, there would not even be a recipient for such a message in this framework.

The constraint of implementability in our model leads to a constrained version of statistical decision theory. Constrained optimal decision functions generally require a PAP. PAPs allow the decision-maker to draw on analyst expertise. Such analyst expertise cannot be used under the alternative mechanism of unilateral specification of decision functions by the decision-maker.

Our model also allows us to derive practical guidance for the design of optimal PAPs. Optimal PAPs lead to constrained-optimal decision functions. We show that the decision-maker's optimal decision function can be implemented by allowing the analyst to choose from a restricted set of decision-functions, and communicating their choice in a PAP. For hypothesis testing, the analyst gets to choose any test that satisfies size control when all data are observed. If a statistic required by the pre-specified test is not reported, then the decision-maker later makes worst-case assumptions about this statistic. The analyst problem, for this mechanism, reduces to a linear programming problem. They have to maximize expected power subject to size control, and subject to the constraints implied by implementability. 
When the set of available statistics is known to the analyst in advance, then the solution to the analyst problem takes the form of a likelihood ratio test.
More generally, we provide an app that allows the analyst to easily solve their optimization problem.

Our model is fairly general in describing the problem of selective reporting by an analyst with conflicting objectives and private expertise.
There are some important considerations, however, which are not reflected in this model.
First, we do not model the potential cost to researchers of documenting complex estimation and testing procedures in the PAP. This is a cost that has been emphasized by critics of the widespread adoption of PAPs \citep{CoffmanNiederle2015,Olken2015,duflo2020praise}. Relatedly, we do not model the cost of communicating complex findings. Such costs likely play an important role in explaining why not all findings are published \citep{Frankel_undated-xn, Andrews2020-cp}.

Second, there are a number of alternative mechanisms that might complement PAPs as tools to limit the adverse effects of conflicting interests and private information.
One such mechanism is adversarial review, where reviewers might request additional statistics to be reported by researchers. Our model does not include a review stage.
Another such mechanism is researcher reputation, and more generally the dynamics of repeated interactions. Our model is a one-shot game, which does not allow for such dynamics.
We hope that future research will elaborate on mechanisms such as these, and the extent to which they might act as a substitute for PAPs.

%% file: Sections/Proofs.tex
\section{Proofs}
\label{sec:proofs}

\subsection*{Implementability}

\begin{proofwithspace}[Proof of \autoref{prop:implementability}]
    We first show that existence of such an $\tilde \act$, which satisfies conditions \eqref{eq:properscoring} and \eqref{eq:monotonicity}, implies implementability. We then show that implementability implies existence of  such an $\tilde \act$.\\

    Assume first that such an $\tilde{\act}$ exists.
    Then, letting the message space be the space of signals $\pi$, and choosing
    $\act(\pi,X_I,I) = \tilde{\act}(\pi,X_I,I)$,
    yields incentive compatibility of $I^* = J, M^* = \pi$:
    For any alternative $\pi,X_J,J$-measurable reporting policy $\tilde{I} \subseteq J$ and message $\tilde M = \pi' $, we have that
    \begin{align*}
      v(\act(M^*,\tilde{I},X_{\tilde{I}})) &\leq v(\act(M^*,I^*,X_{I^*})) \\
      \E[v(\act(\tilde{M},\tilde{I},X_{\tilde{I}}))|\pi]
      &\leq \E[v(\act(\pi',J,X_{J}))|\pi]\\
      &\leq \E[v(\act(\pi,J,X_{J}))|\pi] = \E[v(\act(M^*,I^*,X_{I^*}))|\pi]
    \end{align*}
    The first inequality holds by monotonicity of $\tilde \act$. 
    The first inequality in the second line also holds by monotonicity of $\tilde \act$. 
    The last inequality holds because of the truthful message condition.
      For this choice of $I^*,M^*$, we have $\actreduced(\pi,X_J,J) = \tilde{\act}(\pi,X_J,J)$ almost surely, as desired.\\
  
     Assume now reversely that the reduced-form decision function $\actreduced$ is implementable by a decision rule $\act$, with $\pi,X_J,J$-measurable analyst choices $I^*$ and $\pi$-measurable analyst message $M^* = M^*(\pi)$.
      Define
      $$\tilde \act(\pi, X_J, J) = \max_{I \subseteq  J} \act(M^*(\pi),X_I,I).$$
      Note that $\tilde \act$ is also well-defined for values of $\pi,X_J,J$ outside the joint support of these variables.
      By definition of the reduced form policy, we immediately get
      \begin{align*}
        \actreduced(\pi,X_J,J) =
        \tilde \act(\pi, X_J, J)
      \end{align*}
      almost surely (i.e., on the joint support of $\pi,X_J,J$).

      To see that $\tilde \act(\pi, X_J, J)$ satisfies monotonicity note that the maximum over $I$ can only increase, when it is taken over a larger set of possible values for the set of components $I$.
      To see that $\tilde \act(\pi, X_J, J)$ also satisfies the truthful message condition, note that
      \begin{align*}
        \E[v(\tilde \act(\pi,X_J,J))|\pi]
        &= \E[\max_{I \subseteq J} v(\act(M^*(\pi),X_I,I))|\pi]\\
        &=
        \max_{M \in \mathcal{M}} \E[\max_{I \subseteq J} v(\act(M,X_I,I))|\pi]
        \\
        &\geq
        \E[\max_{I \subseteq J} v(\act(M^*(\pi'),X_{I},I))|\pi]\\
        &=        \E[v(\tilde \act(\pi',X_J,J))|\pi].
      \end{align*}
      The first equality holds given the definition of $\tilde \act$.
      The second equality holds given the definition incentive compatibility for $M^*(\pi)$.
      The following inequality holds since the maximum over $M$ is necessarily weakly larger than the value for any given message $M^*(\pi')$.
      The last equality, finally, again holds given the definition of $\tilde \act$.
      The claim follows.
  \end{proofwithspace}

  \begin{proofwithspace}[Proof of \autoref{prop:implementationalternatives}]
    The first part follows from the arguments in the proof of \autoref{prop:implementability}, where we set   $\act(\pi,X_I,I) = \tilde{\act}(\pi,X_I,I)$.
    Note, in particular, that if a rule is implementable using a $\pi$-measurable message $M^*(\pi)$, then it is also implementable with the signal $\pi$ itself as the message, via the decision rule $\act(\pi,X_I,I) = \act'(M^*(\pi),X_I,I)$.
  
    For the second alternative, implementation using delegation, assume first that $\actreduced$ is implementable by some decision rule $\act$ with message space $\M$.
    Then it is implementable by offering the analyst a choice from $\B = \{(X_I,I) \mapsto \act(M,X_I,I); M \in \M \}$.
    Assume reversely that $\actreduced$ is implementable by the proposed delegation mechanism.
    Then it is implementable by the decision rule $\act(b,X_I,I) = b(I,X_I)$ with message space $\M = \B$.
  \end{proofwithspace}

  \begin{proofwithspace}[Proof of \autoref{prop:proper}]
    The following is based on the proof of Theorem 1 (a generalization of Savage's theorem) in \cite{gneiting2007strictly}.
    A scoring rule is called proper if it satisfies Condition \eqref{eq:properscoring}, the truthful message condition.
    
    We first show that the characterization in the proposition is sufficient for the scoring rule $S$ to be proper.
      Convexity of $G$ and the definition of $S$ based on $G$ immediately imply that $S$ is proper, i.e., that truthful revelation is incentive compatible, since convexity implies
      $$
        S(\pi,\pi) = G(\P_\pi) \geq G(\P_\pi') + \langle G'(\P_{\pi'}, \cdot),  \P_\pi - \P_{\pi'}\rangle = S(\pi',\pi),
      $$
      for any subgradient $G'$.
    
      Reversely, suppose that $S(\pi',\pi)$ is a proper scoring rule. Linearity in $\P_\pi$ holds by definition, since   $S(\pi',\pi)$ is defined, in \eqref{eq:definitionscore}, as an expectation over $\P_\pi$. $S(\pi',\pi)$ is thus, in particular, a convex function of $\P_\pi$.
      $G(\P_\pi) = S(\pi,\pi) = \sup_{\pi'} S(\pi',\pi)$ is an upper envelope of convex functions, and therefore convex itself.
      Furthermore, $S(\pi', \cdot )$ is a subgradient of $G$ at $\pi'$ by definition of proper scoring rules. The claim follows.
    \end{proofwithspace}

    \begin{proofwithspace}[Proof of \autoref{prop:firstbest}]
      Denote by
      \begin{align*}
        \tilde{\act}(\pi,X_J,J) &= \argmax_{A \in \mathcal{A}} \E[u(a,\param)|\pi,X_J,J]
      \end{align*}
      the first-best reduced-form decision rule of the decision-maker.
      Let $\M$ be the set of all signals $\pi$, and choose $\act$ such that $\act(\pi,I,X_I) = \tilde{\act}(\pi,X_I,I)$.
      In this case, $M^*=\pi$ and $I^* = J$ are best responses that implement $\tilde{\act}$.
    \end{proofwithspace}
    
    \begin{proofwithspace}[Proof of \autoref{prop:monotonicity}]
      Suppose first that the monotonicity condition \eqref{eq:setmonotonicity} holds. 
        Then $\act(X_I,I) = \tilde{\act}(X_I,I)$ yields incentive compatibility of $I^* = J$, since for any alternative $\pi,X_J,J$-measurable reporting policy $\tilde{I} \subseteq J$  we have that
        \[
         v(\act(\tilde{I},X_{\tilde{I}})) \leq  v(\act(I^*,X_{I^*})) .
        \]
        by monotonicity of $\act$.
        For this choice of $I^*$, $\actreduced(\pi,X_J,J) = \tilde{\act}(X_J,J)$ almost surely, as desired.
    
       Conversely, consider an arbitrary decision function $\actreduced$ that is implementable by a decision rule $\act$ and $\pi,X_J,J$-measurable analyst choice $I^*$.
        Since $I^*$ is a best-response of the analyst to this decision function $\act$, it follows that the corresponding reduced form decision function satisfies
        $$
          \actreduced(\pi,X_J,J) =
          \act(X_{I^*},I^*)
          =
          \max_{I \subseteq  J} \act(X_I,I)
        $$
        almost surely.
        The right-hand side does not depend on $\pi$, and the maximum (weakly) increases whenever the maximum is taken over a larger set of possible values for $I$.
        The monotonicity condition \eqref{eq:setmonotonicity} follows for $\tilde{\act}(X_J,J) = \max_{I \subseteq  J} \act(X_{I},I)$, which is defined for arbitrary $J$.
    \end{proofwithspace}

\subsection*{Hypothesis testing}

\begin{proofwithspace}[Proof of \autoref{prop:delegation}: ]
  The mechanism described in \autoref{prop:delegation} corresponds to the second characterization of implementability in \autoref{prop:implementationalternatives}.
  Define $\tilde{\B}$ as the set of functions $b$ of the form
  $$b(X_J,J) = \inf_{X';\;X'_J=X_J} t(X'),$$
  for some full-data tests $t: \X \rightarrow [0,1]$ satisfying size control, $\sup_{\param \in \Theta_0} \E[t(X) | \param] \leq \alpha$.
  This $\tilde{\B}$ is the set of decision functions from which the analyst can effectively choose at the pre-analysis stage.
  
  For any such $b$,
  monotonicity of $b(X_J,J)$ is immediate.
  Monotonicity of $b$ and size control of $t$ implies,
  together with $X|\param,\pi,J \stackrel{d}{=} X|\param$ from \autoref{assumption:setup}, that
  $$\E[b(X_J,J)| \param,\pi,J] \leq \E[t(X)| \param,\pi,J] = \E[t(X)| \param] \leq \alpha,$$
  for all $\param \in \Theta_0$, so that $b$ satisfies size control.  

  It remains to show that the $b$ chosen by the analyst has maximal expected power among all decision functions satisfying size control and monotonicity.
  Since the analyst aims to maximize expected power, it suffices to show that 
  for any $\tilde b$ which satisfies size control and monotonicity, the set 
  $\tilde{\B}$ contains a decision function $b$ with power at least as high as that for $\tilde b$.

  To see that this is the case, take any $\tilde b$ satisfying size control and monotonicity. Define $t(X) = \tilde b(X, \{1,\ldots,k\})$, and define $b(X_J,J) = \inf_{X';\;X'_J=X_J} t(X')$.
  Then $b(X_J,J) \geq \tilde b(X_J,J)$ for all $X_J,J$, and $b \in \tilde{\B}$.
  In particular, expected power for $b$ is at least as high as for $\tilde b$.
  The claim follows.
\end{proofwithspace}

\begin{proofwithspace}[Proof of \autoref{prop:knownJ}:]
  Let $\tilde b$ be some solution of the analyst's problem
  Define
  $$
    b(X_J, J) = 
    \begin{cases}
     \tilde b(X_{J'}, J') & J'\subseteq J\\
     0 & \text{else}.
   \end{cases}
  $$
  Then $0 \leq b(X_J, J) \leq \tilde b(X_J, J)$ for all $X$ and $J$, and $b$ satisfies all the constraints if $\tilde b$ does. Furthermore, expected power for $b$ is the same as expected power of $\tilde b$, since the two functions are identical on the support of $P_\pi$. Therefore $b$ is a solution of the analyst's problem.
 
  The second claim follows from an application of the Neyman--Pearson Lemma (cf. Theorem 3.2.1 in \citealt{lehmann2006testing}) to the point null hypothesis $\P_{\param_0}$ and the point alternative $\P_{\pi}$.
 \end{proofwithspace}

To prove \autoref{prop:extremal}, note first that an element of $\mathcal B$ is extremal if and only if there exists no function $\Delta = \Delta(X_J,J)$, where $\Delta \not\equiv 0$, such that both $b + \Delta$ and $b - \Delta$ lies in $\mathcal B$.
\begin{lemma}
  \label{lemma:extremal}
  Suppose that $b \in \mathcal B$.
  Then $b + \Delta\in \mathcal B$ and $b - \Delta\in \mathcal B$ if and only if the following conditions hold:
  \bal
    \int  \Delta(X,K) d\P_{\param_0}(X) &= 0 \label{eq:sizeDelta}\\
    |\Delta(X_J,J)| &\leq \min(b(X_J,J),1-b(X_J,J)) & \forall\; J,X\label{eq:supportDelta}\\
    |\Delta(X_J,J) - \Delta(X,K)| &\leq   b(X,K) - b(X_J,J) & \forall\; J,X.\label{eq:monotonicityDelta}
  \eal
\end{lemma}

\begin{proofwithspace}[Proof of \autoref{lemma:extremal}:]
  Immediate. Each of the three conditions corresponds to one of the conditions defining $\mathcal B$ (size control, support, and monotonicity).
\end{proofwithspace}

\begin{proofwithspace}[Proof of \autoref{prop:extremal}:]
  The first part of the proposition is immediate from our preceding discussion; we prove the characterization of extremal points.
   We first show that the stated conditions are sufficient for $b$ to be extremal.\\
   Suppose $\Delta$ satisfies the conditions of \autoref{lemma:extremal}, and $b$ satisfies the conditions of this proposition. We need to show that $\Delta \equiv 0$.

   \begin{enumerate}
    \item By condition \eqref{eq:supportDelta}, $\Delta(X,K) = 0$ for all $X$ such that $b(X,K) \in \{0,1\}$.
    
    \item If there exists no $X$ such that $b(X,K) = q$, it follows that $\Delta(X,K) = 0$ for all $X$.
    
    \item If there exists only one $X$ such that $b(X,K) = q$, we denote $\Delta(X,K) = \delta$.

    If there exist two points $X \neq X'$ such that $b(X,K) = b(X',K) = q$, then by assumption there is also some $J$ such that $b(X,K) = b(X',K) =b(X_J,J) =b(X'_J,J)= q$ and $X_J = X'_J$. Condition \eqref{eq:monotonicityDelta} then implies $\Delta(X,K) = \Delta(X_J,J) = \Delta(X',K)$. $\Delta(X,K)$ is therefore constant for all $X$ such that $b(X,K) = q$. Write $\Delta(X,K) = \delta$ for such values of $X$. 
    
    It follows that
    $\int  \Delta(X,K) d\P_{\param_0}(X) = \delta  \cdot \P_{\param_0}(b(X,K) = q).$  
    
    \item Condition \eqref{eq:sizeDelta}, in combination with $\P_{\param_0}(b(X,K) = q)>0$ if there exists any $X$ such that $b(X,K) = q$, then implies $\delta = 0$.
  
    \item We have thus shown that $\Delta(X,K) = 0$ for all $X$.
    Condition \eqref{eq:monotonicityDelta}, in combination with our assumption that $b(X_J, J) = \inf_{X':\; X'_J=X_J} b(X', K)$, then implies $\Delta(X_J,J) = 0$ for all $X,J$. The claim follows.
   \end{enumerate}

   We now show the reverse claim, that any extremal point of $\mathcal B$ needs to satisfy these conditions. If any of these conditions is violated, we can construct a $\Delta \not\equiv 0$ which satisfies the conditions of \autoref{lemma:extremal}.

   \begin{enumerate}
    \item Suppose first that there are two points $X,X'$ such that $0<q_1 =b(X,K) < b(X',K) = q_2<1$, so that the first condition of the proposition is violated.
    Let $q_0<q_1 < q_2 < q_3$ be four adjacent points in the range of $b(X,K)$.\footnote{This is the only point in the proof where we use that $b(X,K)$ has finite range.}
    Denote $p_1 = \P_{\param_0}(b(X,K) = q_1)$ and $p_2 = \P_{\param_0}(b(X,K) = q_2)$, and set    
    \bals
      \epsilon &= \min(q_1-q_0,q_2-q_1,q_3-q_2),\\
      \rho_1 &= \begin{cases}
        1 & \text{if } p_1 = p_2 = 0\\
        p_2 & \text{else}
      \end{cases},&
      \rho_2 &= \begin{cases}
        1 & \text{if } p_1 = p_2 = 0\\
        p_1 & \text{else}
      \end{cases}.
    \eals
    Define
    $$
     \Delta(X_J,J) =
       \begin{cases}
         \epsilon  \cdot \rho_1 &\text{if } b(X_J,J) = q_1\\
         - \epsilon \cdot \rho_2 &\text{if } b(X_J,J) = q_2\\
         0 &\text{else}.
       \end{cases}
     $$
     This $\Delta $ satisfies the conditions of \autoref{lemma:extremal}.

    \item Suppose next that the first condition of the proposition holds, and there exists $X'$ such that  $0< b(X',K) = q <1$, but $\P_{\param_0}(b(X,K) = q) = 0$, so that the second condition of the proposition is violated.
    Define
    $$
    \Delta(X_J,J) =\begin{cases}
       \min(q, 1-q) &\text{if } b(X_J,J) = q\\
       0 &\text{else}.
     \end{cases}
    $$
    This $\Delta $ satisfies the conditions of \autoref{lemma:extremal}.
     
     \item Suppose lastly that the first two conditions of the proposition hold, but that the third condition of this proposition is violated.
     In that case there must be two points $X' \neq X''$ such that $b(X',K) = b(X'',K) = q$, and we have that $b(X'_J,J) = 0$ for all $J$ such that $X''_J = X'_J$.

Denote $p_1 = \P_{\param_0}(X')$ and $p_2 = \P_{\param_0}(X'')$, and set
    \bals
      \epsilon &= \min(q,1-q),\\
      \rho_1 &= \begin{cases}
        1 & \text{if } p_1 = p_2 = 0\\
        p_2 & \text{else}
      \end{cases},&
      \rho_2 &= \begin{cases}
        1 & \text{if } p_1 = p_2 = 0\\
        p_1 & \text{else}
      \end{cases}.
    \eals
     Define
     $$
     \Delta(X_J,J) =
       \begin{cases}
         \epsilon \cdot \rho_1 &\text{if } J=K, X=X'\\
         -\epsilon  \cdot \rho_2 &\text{if } J=K, X=X''\\
          0 &\text{if }J=K, X\neq X', X''\\
         \Delta(X,K) &\text{if } J\neq K,b(X_J,J) = b(X,K)=q\\
         0 & \text{else}.
       \end{cases}
     $$
     The penultimate line is well-defined since there is at most one such $X$ (among $X'$ and $X''$) for any given $X_J$, $J$, such that $b(X_J,J) = b(X,K) = q$, given our assumptions.
     This $\Delta $ once again satisfies the conditions of \autoref{lemma:extremal}.
   \end{enumerate}
\end{proofwithspace}

\begin{proofwithspace}[Proof of \autoref{prop:rationalizing}:]
  We construct a prior $\P_\pi(X_J,J)$ such that $\P_\pi(J = K) = 1$, and such that $b$ is optimal within the set of functions $b$ that satisfy size control and the support condition.
  It then follows that $b$ is also optimal within the smaller set $\mathcal B$.

We can define $\P_\pi$ as follows:
  \bals
    d\P_\pi(X_J,J) &= 
    \begin{cases}
      0 & \text{if } J\neq K\\
      d\P_{\param_0}(X,K)  \cdot (2-\alpha) & \text{if }b(X,K) = 1, J=K\\
      d\P_{\param_0}(X,K)  \cdot (1-\alpha) & \text{if }b(X,K) =0, J=K
    \end{cases}
  \eals
  By assumption size control is binding, $\P_{\param_0}(b(X,K) = 1) = \alpha$. This implies that $d\P_\pi(X_J,J)$ integrates to $1$.
  Furthermore, a simple Lagrangian calculation shows that $b$ is optimal for the problem of maximizing $  \int b(X_K,J) d\P_\pi(X_J,J)$ subject to the support condition $b \in [0,1]$, and subject to the size constraint.
\end{proofwithspace}

%% file: Sections/Simple.tex
\section{Case study continued: Simple PAPs}
\label{sec:casestudy_simple}

\begin{figure}[h]
    \centering

    \caption{Optimal simple pre-specified rejection regions for arm-specific cutoffs}

    \label{fig:illustration_simple}

    \begin{subfigure}{.45 \linewidth}
        \centering
        \caption{Known availability}
        \label{fig:illustration0_simple}
        \includegraphics[width=\linewidth]{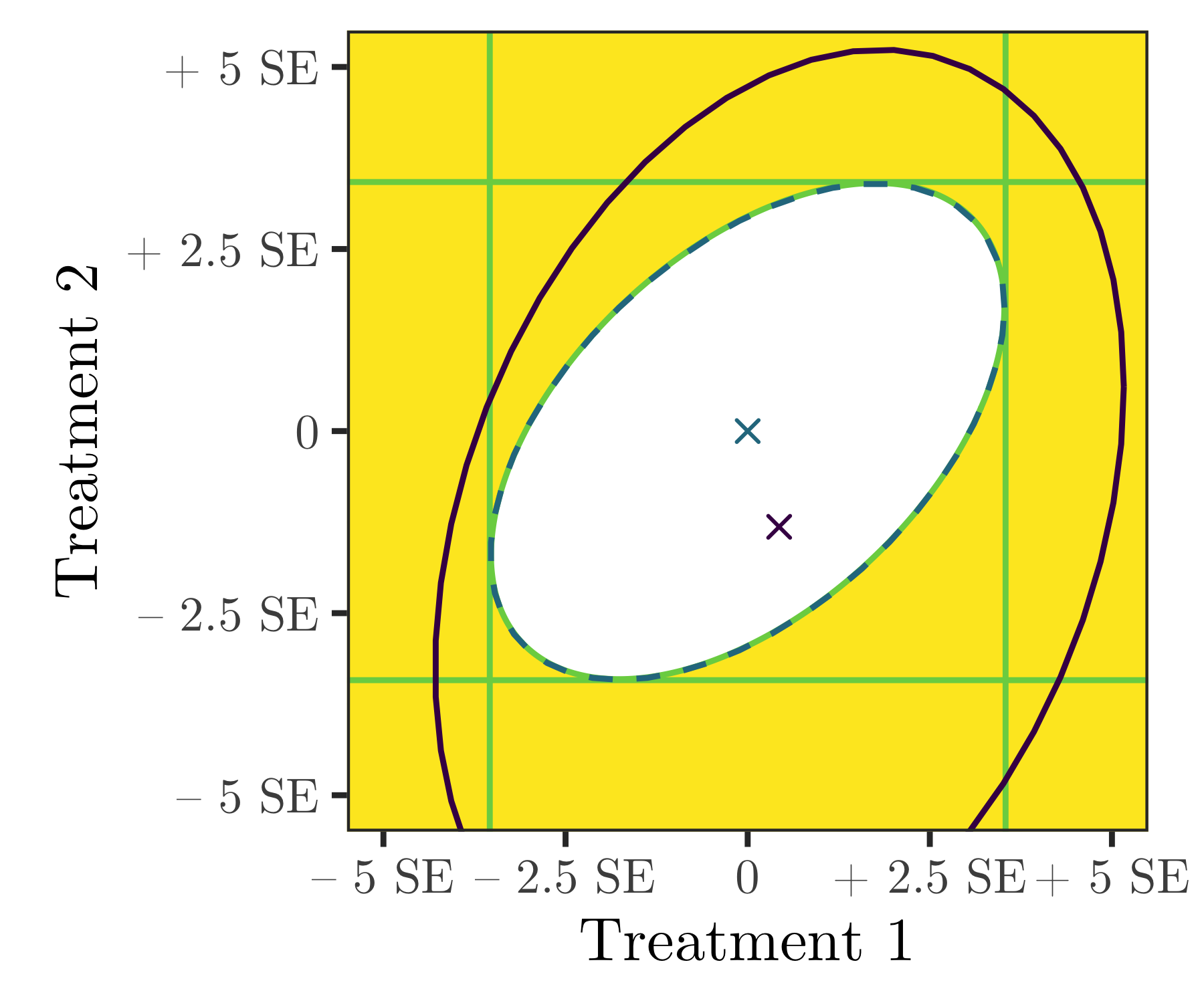}
    \end{subfigure}%
    \hfill
    \begin{subfigure}{.45 \linewidth}
        \centering
        \caption{Uncertain availability}
        \label{fig:illustration1_simple}
        \includegraphics[width=\linewidth]{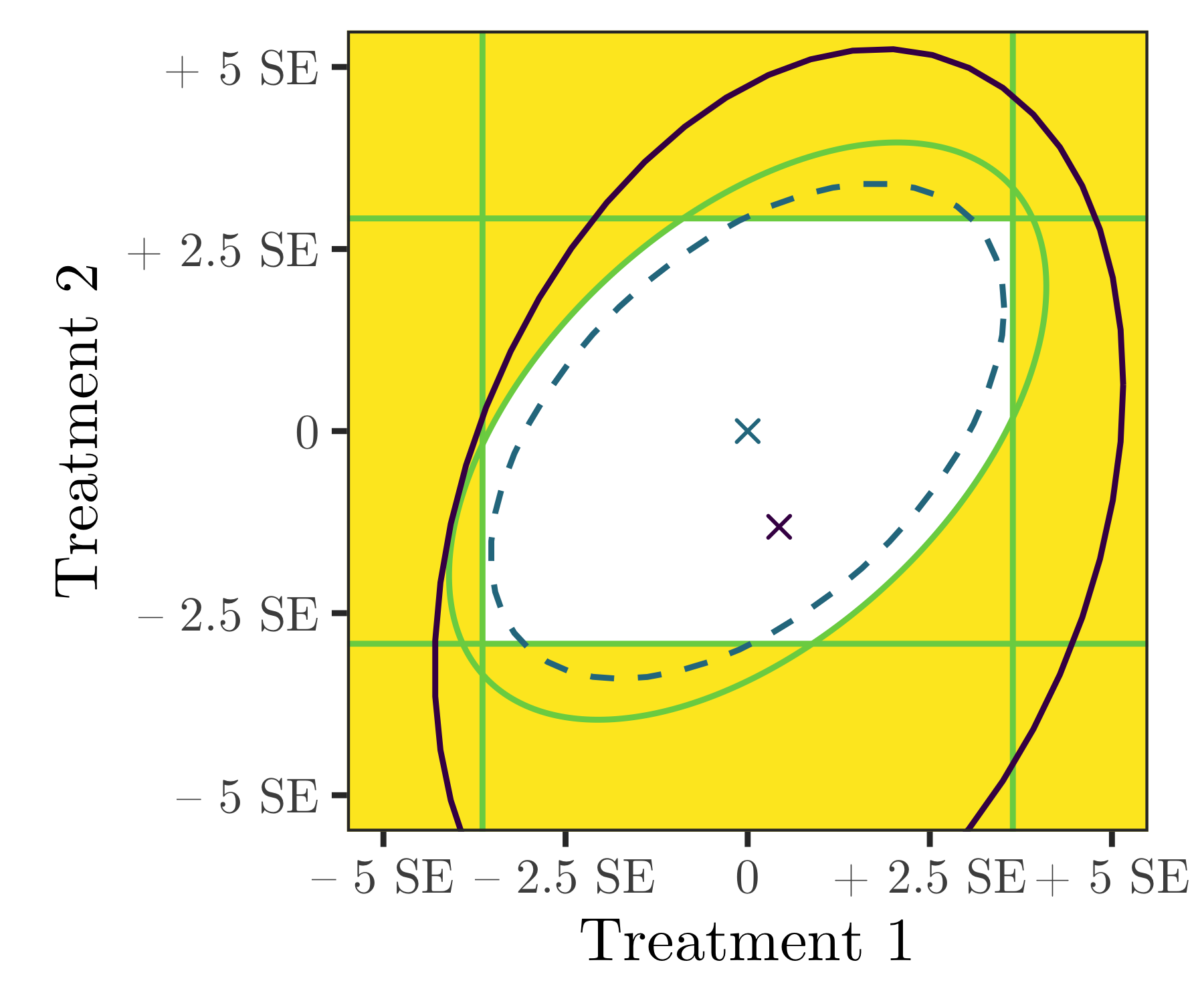}
    \end{subfigure}
    
\end{figure}

In \autoref{sec:casestudies}, we reported first-best optimal PAPs, as well as optimal simple PAPs that can be represented as Wald tests for pre-specified subsets $I$.
In this section, we consider the alternative restriction to PAPs where the set of components $I$ that will be submitted does not have to be pre-specified, but the analyst needs to pre-specify different thresholds $c_I$ for Wald tests for different sets $I$.
Optimal simple tests of this form are reported in \autoref{fig:illustration_simple}.
When the analyst knows that arms 1 and 2 are available, the optimal such test is the same as the simple test from \autoref{fig:illustration0_trivial}.
If, however, the analyst is uncertain about which arms will be available, then the optimal subset-specific thresholds are non-trivial.
The resulting test in \autoref{fig:illustration1_simple} rejects if either the t-statistic for arm 1 exceeds $3.64$, or the t-statistic for arm 2 exceeds $2.92$, or the Wald statistic for arms 1 and 2 exceeds $2.85^2$.

This test approximates the optimal test from \autoref{fig:illustration1_complex} well, with only a small loss of power from 32\% to 31\% due to the simplicity restriction.
Reporting such data-specific thresholds may represent a practical way of committing to effective tests without communicating overly complex rejection regions.
